\documentclass[10pt,letterpaper,twocolumn]{article}
\usepackage{sty/usenix}
\usepackage[square,comma,numbers,sort&compress]{natbib}

\usepackage{silence}
\usepackage[hyphens]{url}
\usepackage[breaklinks]{hyperref}
\usepackage[usenames,dvipsnames]{xcolor}
\hypersetup{citecolor=blue,linkcolor=blue}
\usepackage{amsmath,amsopn,amssymb}
\usepackage{subfigure}
\usepackage{endnotes,microtype,xspace,graphicx,fancyvrb,multirow}
\usepackage{booktabs}
\usepackage{array,underscore,relsize}
\usepackage[T1]{fontenc}
\usepackage{times}
\usepackage{fancyhdr,lastpage}
\usepackage{enumitem}
\usepackage[labelfont=bf,font=small,skip=5pt]{caption}
\usepackage{fp}
\usepackage{siunitx}

\usepackage{balance}

\usepackage{tikz}
\usetikzlibrary{calc,positioning,arrows, shadows}
\usepackage{pgf-umlsd}
\usepackage{comment}

\newcommand{\sys}{\mbox{\textsc{SPX}}\xspace}

\newenvironment{tightitemize}
 {\begin{list}{$\bullet$}{
                \setlength{\leftmargin}{10pt}
		\setlength{\itemsep}{0pt}
		\setlength{\parsep}{0pt}
		\setlength{\topsep}{0pt}
		\setlength{\parskip}{0pt}
		}
 }
 {\end{list}}

\newcounter{tecounter}
\newenvironment{tightenumerate}
 {\begin{list}{\arabic{tecounter}.}{
                \usecounter{tecounter}
                \setlength{\leftmargin}{10pt}
		\setlength{\itemsep}{0pt}
		\setlength{\parsep}{0pt}
		\setlength{\topsep}{0pt}
		\setlength{\parskip}{0pt}
		}
 }
 {\end{list}}

\newcommand{\cc}[1]{\mbox{\smaller[0.5]\texttt{#1}}}

\fvset{fontsize=\scriptsize,xleftmargin=8pt,numbers=left,numbersep=5pt}

\input{code/fmt}

\def\Snospace~{\S{}}

\if 0

\fi

\newif\ifdraft\drafttrue
\newif\ifnotes\notestrue
\ifdraft\else\notesfalse\fi

\input{glyphtounicode}
\newcolumntype{R}[1]{>{\raggedleft\let\newline\\\arraybackslash\hspace{0pt}}p{#1}}

\newcommand{\squishlist}{
\begin{itemize}[noitemsep,nolistsep]
  \setlength{\itemsep}{-0pt}
}
\newcommand{\squishend}{
  \end{itemize}
}

\usepackage{tikz}

\usepackage{xstring}
\newcommand{\PP}[1]{
\vspace{2px}
\noindent{\bf \IfEndWith{#1}{.}{#1}{#1.}}
}

\begin{document}

\title{SPX: Preserving End-to-End Security for Edge Computing}

\author{
 {\rm Ketan Bhardwaj, Ming-Wei Shih, Ada Gavrilovska, Taesoo Kim, Chengyu Song$^\dagger$}
\\
\emph{Georgia Institute of Technology}\\
\emph{$^\dagger$ University of California, Riverside}
}

\date{}
\maketitle

\begin{abstract}
Beyond point solutions, the vision of edge computing is to
enable web services to deploy their edge functions in a multi-tenant
infrastructure present at the edge of mobile networks. However,
edge functions can be rendered useless because of one critical
issue: Web services are delivered over
end-to-end encrypted connections, so edge functions cannot
operate on encrypted traffic without compromising security or
degrading performance. Any solution to this problem must interoperate
with existing protocols like TLS, as well as with new emerging security protocols
for client and IoT devices. The edge functions must remain invisible to
client-side endpoints but may require explicit control from their service-side
web services. Finally, a solution must operate within overhead 
margins which do not obviate the benefits of the edge. 

To address this problem, this paper presents \sys -- a solution for
edge-ready and end-to-end secure protocol extensions, 
%, \sys, 
which can efficiently maintain
end-to-edge-to-end ($E^3$) security semantics.
Using our \sys prototype, we allow edge functions to operate on encrypted
traffic, while ensuring that security semantics of secure protocols
still hold. \sys uses Intel SGX to bind the communication channel with
remote attestation and to provide a solution that not only
defends against potential attacks but also results in low
performance overheads, and neither mandates any changes on the end-user
side nor breaks interoperability with existing protocols.
\end{abstract}

\section{Introduction}
\label{sec:intro}
%Emerging applications enabled by powerful end-user devices and 5G technologies
%create a demand for both reduced access latencies for web services~\cite{gsma:5greport}
%and a dramatic increase in backhaul network capacity~\cite{visualnetworking, ctia}.
%In response to these demands, edge computing---the use of computational resources 
%closer to end devices at the edge of a network to run edge functions~\cite{airbox}
%---has emerged as a solution approach. Going beyond point solutions, 
%researchers~\cite{airbox, mdc, cloudlet} and industry initiatives~\cite{etsi,oec} 
%are pointing towards a vision of edge computing where web services deploy their 
%edge functions in a multi-tenant edge infrastructure. 

Edge computing promises to support low latency web services by moving them
fully or partially (i.e., as application-specific edge
functions~\cite{airbox}) to 
edge computing infrastructure positioned comparatively closer to the data
sources. Edge 
computing infrastructure deployed   
by mobile network operators~\cite{att-anouncement,vodafone-announcement} %or by various emerging enterprise 
%edge-cloud vendors~\cite{edge-cloud1, edge-cloud2, google-globalcache, aws:cloudfront} 
%\ada{enterprise is not clear it's at least} 
%\ketan{I am not referring to google/amazon but several other like conniva, etc. 
%We can leave it out completely and make the pitch around MNOs only. We say that 
%end users will interface with them as opposed to any enterprise edge clouds which 
%can continue to operate with existing threat models but MNOs driven has a different 
%threat model as others won't get the latency benefits to end users. That also ties 
%in with our story of no changes to end users. Thoughts?}
has
at least two orders of magnitude higher degree of distribution 
compared to existing CDNs and cloud datacenters\footnote{100,000s of cellular base stations vs.~1000s of CDNs or 10s of geo-distributed cloud~\cite{fcc,cloudfront-edge,google-map}.}. 
The high degree of distribution exponentially increases the attack surface, 
making edge computing more vulnerable, and therefore, less secure. To guarantee security
properties like confidentiality and integrity for their users, most web services 
employ end-to-end encryption (70\% over TLS~\cite{sandvine}). More
recently, due to the high
TLS overhead for certain use cases like IoT~\cite{ciscosiot} and 
messaging~\cite{whatsapp}, newer secure protocols are being used.
To carry out any useful processing, edge functions require 
access to end users' encrypted traffic in decrypted form, at the 
more vulnerable edge infrastructure. 

Summarizing, edge computing raises a pressing need for a solution
to securely operate on encrypted traffic, in a manner that provides
the {\em performance} benefits of the edge in terms of low latency,
is 
{\em secure} considering existing security semantics and threat models, and
is {\em interoperable} with the new use cases and security
protocols relevant at the edge. 
%\ada{I want to put the requirements,
% and then say why current solutions are not adequate. } This makes sense !

%\noindent{\bf Limitations of prior work. } 
%However, to carry out any useful processing, edge functions require 
%access to end users' encrypted traffic in decrypted form at the 
%more vulnerable edge infrastructure. 
The existing approaches to making
something like this possible are designed for different operating
scenarios and assumed usage models, making them inadequate for edge
computing. 
%
%The existing approaches to making
%something like this possible are inadequate. Their inadequacy stems from 
%either their target operating scenarios, their assumed usage model or their
%design, which make them unsuitable for edge computing
%settings. 
For instance, edge computing applications are posed as the first 
point of contact for low-end devices (e.g., in IoT deployments) with
Internet-based web services. They may employ different security
protocols (other than TLS) 
among them. 
To be practically deployable, 
it is critical for edge computing applications 
and the solutions they adopt 
to be transparent 
to those devices, and not to require device-side changes or active
participation. 
%, in order to make them practically deployable.  
%\ada{This is no longer the sole crux of the problem, because of the
%  other work; highlight low-end devices and different protocols?. }
%\ketan{Tried to address these.}
Existing techniques fail to address these requirements, or introduce
prohibitive performance or security vulnerabilities, as summarized
below. 

%Prior work claims that this can achieved in several ways. 
One approach is to use advanced encryption schemes such as
homomorphic~\cite{gentry} or searchable~\cite{blindbox} encryption, 
which allow edge functions to carry out some operations on encrypted data.
However, the state-of-the-art implementations of these techniques have
very high performance  
overheads that overshadow any latency benefit gained from using edge computing,
ruling out their use. 

Another approach is to leverage recent solutions designed 
for network middleboxes that split secure sessions~\cite{sslsplit,mctls} 
to inspect encrypted traffic. These solutions have several shortcomings. 
%First, mcTLS~\cite{mctls} is based on using different keys for different
%elements of the exchanged traffic, and is essentially a new protocol;
%mandating a new protocol for all devices posed to interact with the edge is not 
%practical and is of limited use. More importantly, splitting a connection~\cite{sslsplit} 
%at the edge requires that  root certificates are installed at the vulnerable edge
%locations, so as to enable edge functions to impersonate web services to clients, and 
%clients to the web services. 
The most severe is their susceptibility to Iago attacks~\cite{iago} by a curious
edge provider employee or to zero-days vulnerabilities such as Heartbleed~\cite{heartbleed} at the edge infrastructure.
Using a shielded execution environment (SEE), such as Intel SGX~\cite{sgxref1,sgxref2}),
as proposed in mbTLS~\cite{mbtls} and SafeBricks~\cite{safebricks}, seems sufficient 
to address this. However, while these can be applied in principle, using them requires 
changes in end user devices. 
mbTLS~\cite{mbtls} claims interoperability with legacy 
devices, however, it only considers TLS, leaving out 
a large number of devices that may not use TLS but still need to be served by edge computing~\cite{ciscosiot}. 
SafeBricks~\cite{safebricks} proposes to use SEE and IPsec to ensure protection 
of encrypted traffic for network functions (NFs) running in the
cloud. It assumes that IPsec capability is introduced in all end user
devices, 
something that depends on how soon IPSec is integrated in those devices. 
Further, a tunnel is needed between each client, edge 
and web service, if we are to make session level end-to-end security guarantees;
this can be too heavy for edge computing use cases, making it practically
difficult to deploy.  
%\ada{why?
%  raluca will review this, we can't just say this without more
%  explanation. }
%\ketan{lets go to details later in the paper}
In summary, although these approaches
are technically apt to address the issues, they may simply be too difficult to apply in edge computing
settings due to their different design goals.
%puts in peril the security properties provided by such protocols 
%because they are not designed for more than two end points for which the security 
%properties provided by such protocols are undefined. 
%Second, the use of mcTLS~\cite{mctls}. 
%However, it requires changes in end user devices to support a new protocol making it
%practically impossible to be used in edge computing scenarios. 
%Even if we assume that, 
%the above two still remain 
%
%
%\ada{In addition, anything else we can say? mbTLS has
%  server-side, so potentially the argument that web services need
%  control of the EFs is not different, and client-side which are
%  optional actually, so not sure the ``overkill'' for edge argument holds. 
%\ada{still thinking... need to go over these again. } 
%\ada{We can now probably just move all these to the related work
%  section, given mbTLS and SafeBricks.}

Furthermore, any system or protocol that relies on an SEE-based remote attestation~\cite{tcg:tpm,intel:sgx} 
to measure the trustworthiness of edge functions and, thereafter, to grant them access to encrypted traffic, 
can be compromised using two type of well known attacks: the time-of-check-to-time-of-use (TOCTTOU)
and the cuckoo attack~\cite{parno:cuckoo} (details in \autoref{sec:threat-model}). 
The root cause of the problem is that the communication with the SEE can be compromised using these attacks.
To maintain the same level of security requires a binding between the SEE and the communication channel.  
However, achieving that securely and transparently is non-trivial, especially for arbitrary security protocols.

\noindent{\bf SPX. }
To achieve that binding and address the abovementioned challenges, we present 
{\bf Secure Protocol Extensions (\sys)} -- a framework for deriving from end-to-end secure
protocols their edge-ready and secure extensions. 
%that enable 
With SPX, edge functions can operate on encrypted traffic over any existing security protocol, in collaboration with
the web services that deploy them, and without violating the end-to-end security semantics of 
the end users' secure channel with the web service. \sys uses hardware-provided
shielded execution environments (SEE) such as Intel SGX and {\em augments} the communication among 
end users and web services via edge functions with an additional communication between the edge and the 
web service {\em at an appropriate step} of the protocol.  
%to achieve its goal without requiring
%any changes on the end users' side. 
With \sys, the web services can enable edge functions to access 
contents of the encrypted traffic while preserving what we refer to as End-to-Edge-to-End 
($E^3$) security semantics (\autoref{sec:des}), equivalent to the same level of
security as the original client device--web service connection.
\sys achieves this without requiring any changes on the end users'
side. 
We prototyped \sys for popular existing protocols: TLS (\autoref{sec:tlx}), 
and protocols developed using the Noise protocol framework~\cite{noise} (\autoref{sec:nsx}) 
to demonstrate its feasibility. The experimental results show
  that the \sys-based versions of these protocols add only modest 12-15\%
  latency overhead. Considering that from end user devices the next
  infrastructure tier (e.g., CDNs) has several factors longer latencies
  than the edge~\cite{telefonica:sec17},
  \sys retains the edge performance benefits, despite these overheads.
%\ada{otherwise we're not saying anything else about maintaining edge performance benefit here. }
%We demonstrate their efficiency in terms of the client's setup connection time and 
%the overhead, which is observed by the client and server on opposite ends of the edge 
%(\autoref{sec:ubench}). Furthermore, we evaluate them with realistic workloads
%(\autoref{sec:real}) used in related research. 

In summary, this paper makes the following research contributions.

\begin{tightitemize}
%\item A practical design approach which preserves security and 
%performance of the existing protocols in edge computing settings 
%using shielded execution environments. 

%\ada{cannot claim this as new contribution. }
\item The notion of $E^3$ security properties, central to enabling edge
  computing, 
made practical through the secure protocol extensions (SPX) approach.

%\ketan{The practicality may be the new contribution but can leave it
%out}
%\ada{possibly. }

\item The SPX design framework for extending secure protocols for the
  edge. \sys can be applied to any protocol. Its design 
considers the limitations of SEEs such as Intel SGX, namely limited
memory, instantiations or processing capabilities, and makes  
recommendations for supporting vectored instructions in next generation shielded 
execution environments to make them more attractive for use in edge computing. 
%\ada{not sure can claim this as new contribution. } \ketan{I think recommendation for hardware advancements in future processors is definitely a contributions.}

\item Concrete SPX implementation for TLS and any protocol developed
  using the Noise protocol 
framework, and evaluation of their performance impact on latency, the core reason driving edge computing.

\end{tightitemize}

\section{Motivation}
\label{sec:moti}

We start with an example scenario to clarify the motivation for the development of
SPX.

\PP{Motivating example scenario}
Consider a company like ADT security. It decides to deploy its edge function 
that analyzes data from sensors deployed at customers' homes 
to ensure timely detection of unusual activity, 
such as house break-ins or fire breakouts, in a cost effective
manner. The benefit expected by ADT is improved level of service, due
to more timely incident detection, and reduced operating costs,  
resulting from reduced backhaul bandwidth usage relative to having to deliver 
all traffic from the sensors to ADT's backend server. 
Today, the various sensors in a single home connect to an ADT hub device, deployed
at the home, which 
in turn interacts with their cloud based web service using TLS or other end-to-end secure protocols.
We posit that to lower the costs of such services with the increasing number 
of sensors and other IoT devices in the homes (e.g, smart switches, appliances,
etc.), and to make them more manageable (e.g., in light of the recent incidents~\cite{mirai,dyn} caused due to vulnerabilities in 
devices such as routers, set top boxes, etc.), 
these hubs will be replaced with 
edge functions running on a third-party multi-tenant edge
infrastructure, such as what will be offered by mobile network operators~\cite{openfog,etsi}.
In that case, the individual sensors in homes will have to use TLS, or another end-to-end
secure protocol, to connect to the web service, and these connections
may be transparently routed to a nearby edge location, for ADT to
achieve the expected benefits. 
In this example, a connection must be established between the sensors
and ADT's backend server,  
while the edge function is able to look at and operate on the encrypted traffic to analyze it.

However, when a sensor connects to an ADT edge function, 
it is desired that the connection terminates with the same security as if it were terminating at an ADT 
web service. This is different from today's scenario where content delivery networks (acting in reverse) 
are used to aggregate the traffic from hubs or devices. Today, their secure connections are terminated at CDNs 
where root certificates are installed and security guarantees rely on
the physical security of
the CDN nodes via business agreements and the TLS protocol, as shown in~\autoref{fig:interaction}. 
In edge computing, this may not be acceptable due to the huge attack surface.
%In this example, a connection must be setup between the sensors and ADT's backend server 
%while the edge function is able to look and operate on the encrypted traffic to analyze it.  
Further, not all sensors may be capable of supporting TLS due to the
additional RTTs, the size of the certificates,
and battery constraints~\cite{cloud-flare-blog}. %\ada{what do you want
%  to say here? However, this example does not fully represent the
%risk of compromised edge functions.}
%\ketan{was joining line for the section around security risk in edge computing ... not needed any more}

From the example, we derive the following requirements for a practical
solution for the edge:
\begin{tightitemize}
\item {\em Support unmodified end user devices:} It is critical to not require changes 
in end user devices or the protocols they use. At the same time, edge
functions must be enabled to securely operate on
encrypted traffic while maintaining security guarantees;
\item {\em Consider vulnerable infrastructure:} With a huge attack
  surface, the trustworthiness of the 
privileged software at the edge is questionable especially in multi-tenant settings, as 
highlighted by the NSA-linked Cisco middlebox zero day vulnerability~\cite{nsa-cisco-zeroday} 
incident.
\item {\em Pose minimum performance overhead:} A high performance
  overhead for mechanisms to ensure  
confidentiality and integrity of encrypted user traffic can absolve the latency benefits 
expected from the edge functions
\end{tightitemize}
However, these requirements do not appear to be addressed in leading industry security specifications.
%\PP{Security risk of compromised edge functions} The magnitude of security risk of an edge function 
%being compromised is huge. First, it would be difficult to detect a compromised edge function 
%as it would register itself as valid edge function. Second, it can leak data and sensitive user 
%information of users. Finally in multi-tenant settings, a compromised edge function not only 
%poses risk to the web service that deployed it but can potentially compromises other co-located 
%edge functions and thus, providing a back door to the cloud based web services which deploy 
%those other edge functions. So, with edge computing's inherent vulnerability due to huge attack 
%surface and prevalence advanced persistent threats such as zero-days, malicious insiders etc., 
%even relying on privileged software (OS or hypervisors) for security is unreasonable. 
%However, this is something being not appropriately addressed
%by leading industry specifications.

\PP{Gaps in Industry Specifications} Leading industry bodies~\cite{ngmn,etsi,openfog} 
responsible for defining a secure edge computing ecosystem have suggested directly putting the private key material of the 
certificate into an SEE such as SGX to provide the required security. 
However, this will not suffice. First, a certificate ``kind of'' binds the DNS 
name to a private key, and browsers/apps on end user devices would
only verify the DNS name if it matches the name in 
the certificate, but the edge function may not have the same DNS name. Second and more importantly, although 
we trust SEE to provide isolation, it does not guarantee the software running inside SEE is bug free. This is also 
the main motivation for adding ASLR support in Intel SGX. 
In case of vulnerability such as Heartbleed, the private 
key could be leaked very easily. Moreover, schemes based on time
invalidation of certificates have known
vulnerabilities. Therefore, putting private key
material in the SEE is risky. Recent work on SEE-based middleboxes aim to
address a similar problem, but by focusing on datacenter-centric
assumptions and design constrains, these efforts remain open to
certain types of vulnerabilities~\cite{safebricks}, are limited 
to certain types of devices and protocols~\cite{mbtls}, or their 
impact on the latency overheads is not discussed (see~\autoref{sec:eval}).

\section{Assumptions}
We make the following assumptions about operating scenarios.
First, we assume that backend services are willing to deploy edge
functions on open multi-tenant edge infrastructure. Edge functions
handle end-user requests over secure protocols to provide benefits in
terms of reduced
latency and bandwidth usage.
%At the same time, backend services want end-to-end guarantees to
%preserve the privacy of 
%their users, be informed and/or authorize the edge function to interact with 
%users on their behalf, and to secure their states that lie within edge functions.
Second, we assume that when a client tries to access a web service,
the connection is transparently directed to an edge function of the web service.
The redirection can be DNS-based or S1-based, in enterprise and the mobile 
edge, respectively, or some application level redirection. The mechanism used 
for redirection is orthogonal to the design of \sys and we assume that 
all relevant packets in a flow are redirected to an \sys-enabled edge function. 
Third, we assume that security guarantees provided by existing protocols
in end-to-end communication, such as TLS or Noise-based protocols, still hold. 

Finally, we assume availability of an OS shielded execution environment (SEE) 
whose trust can be established via remote attestation. However, 
we do not make any assumptions around the performance of the available SEE.
Designing an SEE-based solutions has inherent performance overhead 
due to memory encryption, limited or no I/O access in trusted mode, 
limited addressable memory,
% (e.g., Intel's 6th generation processors 
%have a maximum of 128 MB addressable by an enclave), 
and limited number 
of concurrent SEE executions.
% (e.g., Intel's 6th generation processors can 
%support a maximum of 12 enclaves). 
%\ada{I removed these, we'll need space, and we repeat this later
%again. }
In addition, compared to a canonical threading 
model (e.g., pthreads), multithreading within an enclave is not trivial~\cite{sgxref2,sgxref1}. 
Some of 
these limitations are addressed in the newer architectures, but the fact remains 
that a shielded execution in an enclave is and will remain a scarce resource on processors. 
This is important because it obviates some of the trivial solutions. For example, allowing every edge 
function to keep a secure channel with the corresponding web service
open at all times, and subsequently, to  
use that channel to get session keys for all clients, say based on a unique nonce of the
session, has two short comings: First, it could lead to highly inefficient use 
and wastage of SEE resources, and second, it still does not suffice for end-to-end
security semantics.
%below. \ada{there wasn't anything "below"}

%Next, we specify the threat model we considered in
%designing \sys.

\section{Threat Model}
\label{sec:threat-model}
Different from a traditional network threat model, where an adversary usually
is in the middle as an observer and launches attacks only over the traffic on a secure channel, 
we consider a stronger threat model where an adversary has full control over the 
edge infrastructure, including the operating system, except for the shielded execution 
environment (SEE). The adversary can freely monitor, intercept, and forward the data
over the communication channel between the OS and SEE. For example, if an edge function
running in an SEE tries to open a channel to the OS in order to talk to remote clients, 
the adversary who controls the OS also has the full control over that channel. 
The rationale behind considering such a threat model is derived from the following observations:
\begin{tightitemize}
\item Edge computing is a multi-stakeholder and multi-tenant environment with 
different ownership domains, i.e., mobile network operators' edge infrastructure, web 
services's edge functions, and end users' data. 
\item The adversary can be a curious edge infrastructure provider (insider) or
a malicious edge function exploiting a zero-day vulnerability to gain root access.
With that, it can potentially compromise all edge functions and potentially open a 
back door to the cloud-based web service. 
\item Finally, the edge infrastructure is typically deployed in physically insecure locations 
(cellular towers, aggregation points, edge servers, etc.) creating a
huge attack surface of edge computing, 
and protecting it from advanced persistent threats like zero-days,
malicious insiders, etc., 
mandates stronger threat models.
\end{tightitemize}

The above observations around edge computing imply that it warrants stricter security and 
lead us to consider the stronger threat model.
We argue that the existing solutions that
consider a weaker threat model %that is weaker and
are not applicable in edge computing scenarios due to their target use cases or operating assumptions.
The attacks discussed above could lead to a situation where a malicious edge function has privileged access 
and can mount Iago attacks~\cite{iago}, and warrant an SEE-based
  solution.

Furthermore, most of the existing solutions fail to adequately address 
the lack of binding between communication channel and the SEE %adequately 
when using remote attestation. If a communication channel is not carefully bound, two 
general types of attacks are possible:

{\bf TOCTTOU attack:} A communication channel is vulnerable to a TOCTTOU attack if the
attestation is performed \emph{before} establishing the communication channel. Consider the following
scenario:
A web service initializes a remote attestation request to measure the target edge function. Next, 
the malicious edge function with root privileges routes this request to the intended benign edge function. 
The benign edge function responds with an expected measurement and the web service initializes 
another encrypted communication channel to an edge function. 
%To provide confidentiality and 
%integrity (e.g., using the TLS protocol), a 
The malicious edge function routes the new communication 
to itself instead. 

{\bf Cuckoo attack:} A communication channel is vulnerable to a cuckoo attack if the
attestation is performed \emph{after} establishing the communication channel.
When a web service initializes an encrypted channel to an edge
function, the malicious edge function routes the communication to itself. The web service initializes a remote
attestation request over the established channel. A malicious edge function then forwards the attestation request 
to the correct edge function which responds with a correct measurement
to the malicious edge function. The malicious
edge function then forwards the correct measurement to the web service. 

These attacks are possible for two reasons.
First, the malicious edge function has full control of its internal network, so it can arbitrarily 
route network packets to whichever edge function it wants. Second, because there exists a lack of a 
trusted channel to deliver the certificate, %TLS 
the end-to-end protocol
cannot distinguish among malicious and 
benign edge functions, potentially leaking sensitive key material.  However, if we just use a short 
lived Diffie-Hellman (DH) key bound to the SEE attestation, even if the edge function is compromised, the consequence 
is much less severe. However, the SEE performance overhead in doing that has to be within
acceptable limits to make it suitable for edge computing.

Under the threat model, we assume that the adversary cannot mount any attack directly on the SEE. 
We assume the SEE is under a strong protection against
all external software accesses. We do not consider denial of service attacks of edge functions
that can be mounted by the privileged software at edge infrastructure. This is reasonable because
devices can simply fallback to interacting with the web service directly in case edge functions
are under a denial of service attack. We also do not consider side channel attacks on SEE.
\begin{figure}[t]
\centering
\includegraphics[width=\columnwidth]{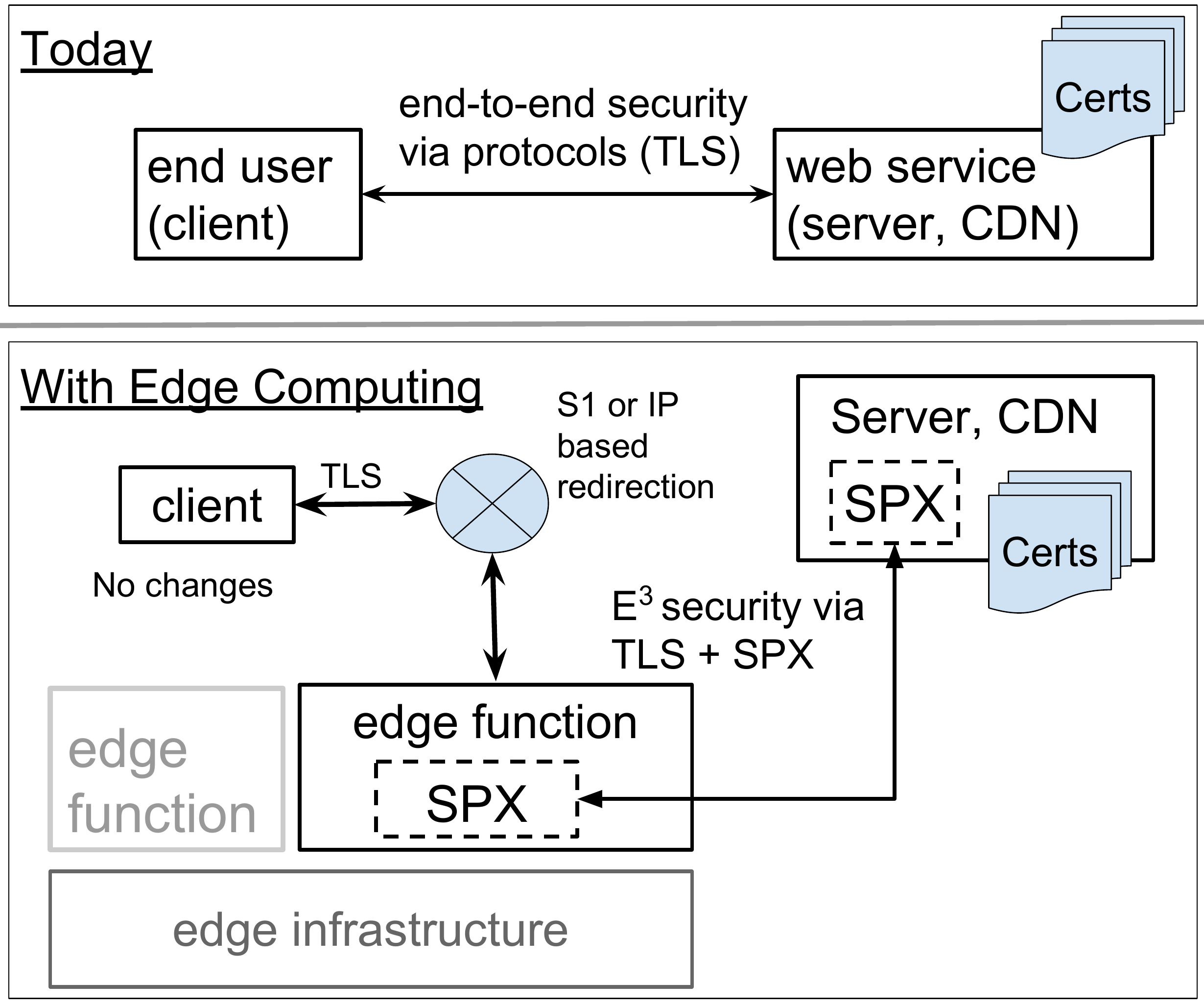}
\caption{Comparing interaction between end users (client) and web services (server) that operate on assumption of end-to-end security using protocols like TLS vs. the interaction between end users client, edge functions and web services which violates assumptions in the protocols rendering security provided by them undefined. With \sys we define it as $E^3$ security.}
\label{fig:interaction}
\end{figure}

\section{\sys Overview}
\label{sec:des}

%\autoref{fig:over} shows the simplified view of how \sys can be realized for any secure protocol,
%using an OS-agnostic communication channel secured by remote attestation provided
%by the TEEs. We next discuss the edge- and server-side details of \sys during different phases
%of a protocol lifetime.

%\begin{figure}[t]
%\centering
%\includegraphics[width=\columnwidth]{fig/overview}
%\caption{Showing the overview of SPX approach.}
%\label{fig:over}
%\end{figure}

The primary goal of any end-to-end protocol like TLS or the
Noise-based protocols is to provide privacy and data integrity between
two communicating applications. These applications are assumed to be
running on two ends, i.e., client and server. However, in case of edge
computing, it is desired that the edge, which is in the middle of the
client and server, is afforded access to their communication so as to
complete performance-enhancing functions.  

\sys enables creation of a secure communication channel between the end user (client) and 
the edge function transparently, with the same end-to-end security semantics as 
when the client directly opens a secure channel with the web service (server), as shown
in~\autoref{fig:interaction}. This is needed to enable edge functions to perform
useful application-specific processing, beyond just
network-level functions.
If edge functions cannot access the encrypted data in the messages
from a client
to a server, edge computing is only of limited use. 
%As a result, a more and useful functionality 
%cannot be implemented with in edge computing settings. 

In principle, it can be argued that terminating connection at an edge function is still 
end-to-end secure by simply splitting TLS. However, it should be noted that in that interaction, one 
end has moved from the web service to the edge function. This termination of 
the secure session at a potentially insecure edge function is not the same. Thereby,
we argue that it does not have same security semantics as terminating
it at secure server.  
We propose \sys with $E^3$ security semantics as a key enabler in edge computing
without compromising the security or usefulness of edge functions.

Concretely, we consider \sys to be satisfying $E^3$ semantics 
 for a given protocol if and only if \sys preserves the end-to-end 
 security semantics for encrypted network traffic transmitted 
 using that protocol, i.e., it preserves confidentiality, 
 data integrity and authentication between client and server, 
 from every other entity in the system (including any privileged
 software on the edge 
 infrastructure such as OS or hypervisor), except for the 
 edge function. 

This is achieved in two steps. First, \sys utilizes SEE to perform the 
encryption so a malicious edge function (even with root access) can only 
see the encrypted network traffic. Second, \sys utilizes an attestation-bound 
handshake protocol to transfer the session key of the supported protocol 
from the server to the edge at an appropriate point in the protocol.
As a result, when establishing the secure channel, the client still 
authenticates using the server's credentials (not deployed at edge functions).
In other words, the edge function is transparent, e.g., during a handshake, the client 
will see and authenticate the edge function using the server's certificate instead 
of the edge provider's certificate. Moreover, the authentication key of the \sys-enabled 
secure protocols (e.g., the private key of the server's certificate)
never leaves the server so the risk of identity theft is lowered and
is the same as in
today's cloud-based web services. Finally, by binding the key exchange between the 
edge function and the server with an SEE-based attestation, \sys is resistant to the 
two kinds of attacks discussed earlier in~\autoref{sec:threat-model}. Achieving this
however requires different design considerations than typical security protocols, 
discussed next.

\section{\sys Design Considerations}
The design considerations for \sys critical for achieving $E^3$
security semantics and retaining edge-based performance can be
described as follows:

\PP{Ensuring security semantics}
To preserve security semantics, the following key elements must be considered:

{\em Channel binding:} The key to prevent the TOCTTOU and cuckoo
attacks (and other attacks as well) at the edge is to bind an attestation with
the corresponding communication channel.
In \sys, this is done by including the ephemeral public key in the
attestation so
% when the server verifies
%the integrity of the edge function it can check 
the server can verify at the same time 
(1) that the edge
function is indeed running inside an SEE,
(2) the integrity of the edge function, and (3) that the ephemeral key pair is indeed generated inside the SEE
so no other entity has access to it.

{\em Channel relaying:} After successful binding, the next important
step is when the server shares the  session state and session key with
edge function, before  
communicating the completion of the secure connection establishment to
the client. This allows an \sys-enabled edge
function to relay the same communication channel used for its own
handshake to the clients. This is important
(i) to ensure that the attested connection is relayed to the client,
hence, establishing its security properties,  
and (ii) requiring additional connections is undesirable from the
perspective of the edge function and server resource usage.

%\ada{... something missing here}
%\ada{cryptic, please rewrite: }

\PP{Preserving performance} 
As noted earlier, latency performance is a key benefit of edge
computing, and any security solution for the edge must preserve this benefit.
Below are more performance oriented design elements that must be
considered. 

{\em Piggybacking.} \sys messages are piggybacked with existing protocol messages as much as possible.
%as opposed to detached messages. 
\sys piggybacked markers trigger \sys operations%function 
%\ada{operation -- trying to avoid using edge functions and \sys
%  functions, the later should be steps or operations. }
at the edge functions
or server. 
By doing this, \sys minimizes the need for extra interactions between
the edge and the 
server,  
which reduces the possibility of man-in-the-middle (MITM)
attacks. This also to maintains at minimum the overhead on 
the edge and server side, and keeps the client side engaged to avoid timeouts in existing protocols. 
%\ada{what does piggybacking have to do with that. can the
%  same not be achieved with standalong messages? doesn't the
%  edge/server runtime still have to detect the \sys marker and
%  dispatch it?}

{\em Protocol replication.} Wherever possible, \sys running in the
edge function replicates the protocol's abstract state at the edge, 
instead of having the edge function maintain all state variables and
compute all crypto functions, which can be resource intensive. 
%\ada{this is not clear, replicate
%  state and the edge as opposed to maintain state at the edge??}
Replicating state allows \sys-enabled edge functions to keep track of the 
protocol state and carry out operations like requesting the session key at
the right time during a handshake via remote attestation, so as to ensure that
even without actually carrying out crypto functions, the channel binding remains 
secure. This allows \sys to 
keep the computational overhead associated with carrying out those functions at 
the edge in check.

{\em Addressing limitations of SEE hardware.} It is important to consider the
limitation of the SEE hardware. For instance, 
(i) the 6th generation Intel processors have only 128 MB of memory
addressable to all the running enclaves, 
 (ii) the number of enclave instantiations running simultaneously is limited,
and (iii) running an SGX thread consumes one of the logical CPUs and makes it unavailable to other
threads. Although these are soft limitations that are further relaxed
in current and future processor generation,  a practical \sys solution
should not be bound by them. 
%be able to overcome them. 
To address this, 
(i) \sys uses a spilling mechanism, i.e., the session look up table
spills over to host memory if the enclave does not have enough
  memory.
%the memory available to enclave is not available. 
However, to ensure $E^3$ guarantees,
\sys uses the SEE  {\em sealing} feature to store session information. To guarantee the
confidentiality and integrity of the sealed information, sealing encrypts the information using
the hardware generated sealing key. By doing this, with a small performance penalty,
\sys-enabled edge functions are only limited by the amount of {\em host} memory as opposed to
memory addressable by the SEE. To mitigate the impact of (ii) \sys
enabled edge functions require 
only one enclave instantiation per edge function as opposed to one enclave instantiation per session.
However, \sys cannot share enclaves with two different edge functions without compromising $E^3$
guarantees, because in that case two edge functions can generate the
same ephemeral key and the same
attestation reports. To mitigate the impact of (iii), we need to make sure that the time required
in binding is kept to a minimum so that the edge function waiting for attestation response,
including session keys, finishes as quickly as possible. This is accomplished by monitoring 
the state of the handshake.

To ensure that these considerations are always met, we defined a set of key operations common
to arbitrary secure protocols, discussed next.

\section{Designing a Generic \sys}
\label{sec:generic}
The following key operations are needed to design an \sys for an arbitrary secure protocol.

{\em Detect ({\bf D}):} An \sys-enabled edge function must be able to detect the client's
session initiation message which is usually transferred in plain text.
% to communicates edge function's intent to use \sys capability.
% On receiving an \sys intent,
% a \sys capable server must communicate its capability to edge function.

{\em Relay ({\bf R}):} Once the edge function detects an initialization of an \sys-enabled protocol,
it relays the initialization message to the server and behaves like a transparent proxy until
the session is established.
It is important that the edge function monitors the entire handshake process so as to extract
relevant protocol state (e.g., encryption suite) that is necessary to talk to the client.

{\em Bind ({\bf B}):} Piggybacking with the relayed traffic,
%the edge function and server create another secure communication channel for key exchange and
%perform mutual authentication that binds an attestation to the channel.
the edge function sends an additional message that binds the attestation with an ephemeral public key
to the server. The ephemeral key pair is generated inside the enclave and will be
deleted after the handshake completes.
%In this step, we use ephemeral Diffie-Hellman as the key exchange method.
This is the step where the two aforementioned attacks are prevented.

{\em Forward ({\bf F}):} To maintain transparency, the edge function must not communicate 
any \sys-related information in the protocol messages exchanged with clients. 
The extra steps in the protocol must be handled at the edge so that clients
receive the same information as if they were setting up a direct connection
with the web service.
% For example, on detecting that the server supports \sys capability or after the attested-binding step finishes,
% edge function must facilitate message transfer in transparent manner.
This function is important to address interoperability and deployment concerns with \sys.

{\em Grant ({\bf G}):} When the server finishes the session establishment and the authentication of
the edge function is successful (by verifying the attestation report),
%(i.e., it is untampered, indeed runs in a TEE, and the key exchange channel is secure)
the server transfers the session key, which is encrypted by the ephemeral public key contained in the
attestation, to the edge function through the key exchange channel.
%to the edge function through the key exchange channel.
On receiving the session key message, an edge function decrypts the
key using the corresponding ephemeral
private key, registers the session, and begins serving the client's request thereafter.

%After successful creation of a secure session, edge function maintains the state of the session
%unless explicitly directed to discard.
{\em Resume ({\bf S}):} \sys may optionally also keep state for the session if a protocol supports 
resume or zero-RTT session setup, e.g., TLS1.3 or Noise protocols with pre-shared keys in IoT devices.

Next, we describe how we applied the above framework to develop \sys for TLS and Noise-based protocols.

\section{Design of TLX = TLS + SPX} %We now describe the TLS protocol operation with \sys:
\label{sec:tlx}

TLS/SSL~\cite{rfc5246} has two types of subprotocols: the handshake protocol and the record protocol.
The handshake protocol is for negotiating security parameters for the record protocol, authenticating two peers, reporting errors, etc.
It is also the key step to prevent MITM attacks.
In particular, by using public key encryption, TLS guarantees that only the two communication peers have access to
the randomly generated session key that is used in the record protocol.
In TLX, we utilize \sys to securely transfer this secret key from the server to the edge function without trusting the edge provider.
\autoref{fig:tlx} shows the TLX protocol with SPX operations marked with same letters as defined in ~\autoref{sec:generic} and is described below:

\begin{figure}[t]
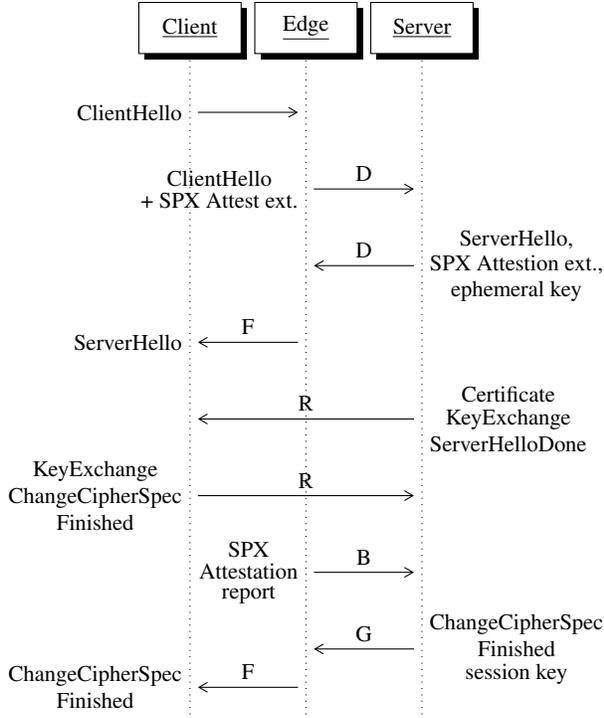

\centering
\tikzset{every picture/.append style={transform shape,scale=0.85}}
\begin{sequencediagram}
\newinst{client}{Client}
\newinst{edge}{Edge}
\newinst{server}{Server}
\mess{client}{}{edge}
\node[anchor = east](ch) at (mess from) {ClientHello};
\postlevel
\mess{edge}{D}{server}
\node[anchor = east](chx) at (mess from) {\shortstack{ClientHello \\+ SPX Attest ext.}};
\postlevel
\mess{server}{D}{edge}
\node[anchor = west](shx) at (mess from) {\shortstack{ServerHello, \\SPX Attestion ext., \\ephemeral key}};
\postlevel
\mess{edge}{F}{client}
\node[anchor = east](sh) at (mess to) {\shortstack{ServerHello}};
\postlevel
\mess{server}{R}{client}
\node[anchor = west](ckx) at (mess from) {\shortstack{Certificate\\ KeyExchange \\ ServerHelloDone}};
\postlevel
\mess{client}{R}{server}
\node[anchor = east](kcf) at (mess from) {\shortstack{KeyExchange \\ChangeCipherSpec \\Finished}};
\postlevel
\mess{edge}{B}{server}
\node[anchor = east](arb) at (mess from) {\shortstack{SPX \\Attestation \\report}};
\postlevel
\mess{server}{G}{edge}
\node[anchor = west](csk) at (mess from) {\shortstack{ChangeCipherSpec \\Finished \\session key }};
\mess{edge}{F}{client}
\node[anchor = east](ccf) at (mess to) {\shortstack{ChangeCipherSpec \\Finished}};
%\mess{client}{}{edge} 
%\mess{edge}{Encrypted messages}{client}
\end{sequencediagram}
\caption{Showing the SPX enabled TLS: TLX handshake protocol.}
\label{fig:tlx}
\end{figure}

A TLX enabled edge function detects a \cc{ClientHello} and \sys adds a request as a TLS extension 
    in the \cc{ClientHello} message which is then forwarded to the server.
Having received the request, the server responds by including a response in its \cc{ServerHello} message,
    again, as a TLS extension.
    The response can be \cc{OK} or \cc{Not Capable}. An \cc{OK}
    message is followed by 
    a challenge (nonce), to guarantee the freshness of the
    attestation and to prevent replay attacks, 
    and a signed ephemeral public key (similar to the \cc{ServerKeyExchange} message), to establish the key exchange channel.
Having received the response from the server, \sys strips the server response from the \cc{ServerHello} message
    and relays the rest of the messages to the client.
\sys validates the certificate, checks whether it is from the trusted server
    (similar to public key pinning~\cite{rfc7469}),
    and verifies the ephemeral key is signed by the public key which corresponds to the certificate
    to prevent MITM attacks~\cite{liang2014https}.
While waiting for the client to respond, \sys generates a fresh pair of ephemeral keys and
    generates an attestation that includes both the public key and the challenge (nonce).
    Then it sends the attestation and the public key to the server together with the client response.
Having received the attestation, the server verifies its
      freshness, correctness, 
    as well as the legitimacy of the public key.
    Once every check clears, the server then sends the session key to the edge function encrypted with
    the exchanged ephemeral key, as well as \cc{ChangeCipherSpec} and \cc{Finished} messages.
    Note that for transparency, all \sys-related messages will not be included when
    calculating the hash of handshake messages.
\sys strips and decrypts the session key and forwards the
      rest of the messages to the client.
    From now on, the edge function can securely communicate with the client using the TLS record
    protocol with agreed cipher spec.
%\end{tightenumerate}
 
The handshake is aborted if either the normal TLS handshake fails or an attestation-related error occurs.

\section{Design of NoiXe = Noise + SPX}
\label{sec:nsx}

{\bf Noise Background:} Noise~\cite{noise} is a framework for crypto
protocols based on the Diffie-Hellman key agreement. A Noise handshake
is described by a simple language consisting of tokens arranged into
message patterns. Message patterns are arranged into handshake
patterns. A handshake pattern specifies the sequential exchange of
messages between an initiator (client) and responder (server), and the
corresponding crypto state transitions that comprise a
handshake. A handshake pattern can be instantiated by crypto functions (DH functions, cipher functions,
and a hash function) to give a concrete Noise protocol. A Noise protocol starts with prologue data and
some patterns require pre-messages to be exchanged between the
parties. 
%In a nutshell, 
Messages
include {\tt e:} ephemeral key, {\tt s:} static key, and {\tt dhee,
  dhse, dhes, dhss:} the sender performs a DH between the
  corresponding local state, remote state and computes the required hashes.  
  Readers are directed to the Noise specification for further details about the protocol framework~\cite{noise}.

\begin{figure}[t]
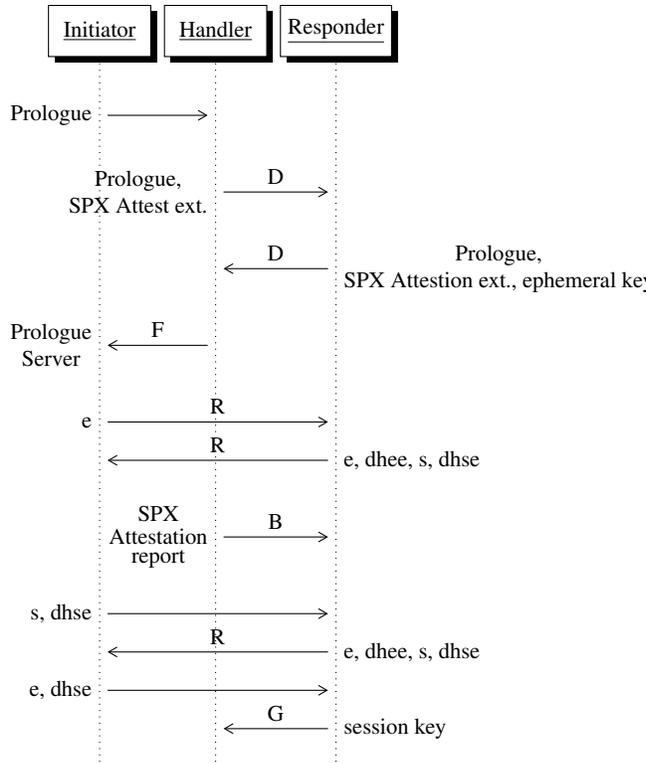

\centering
\tikzset{every picture/.append style={transform shape,scale=0.85}}
\begin{sequencediagram}
\newinst{client}{Initiator}
\newinst{edge}{Handler}
\newinst{server}{Responder}
\mess{client}{}{edge}
\node[anchor = east](plg) at (mess from) {\shortstack{Prologue}};
\postlevel
\mess{edge}{D}{server}
\node[anchor = east](plx) at (mess from) {\shortstack{Prologue, \\SPX Attest ext.}};
\postlevel
\mess{server}{D}{edge}
\node[anchor = west](pls) at (mess from) {\shortstack{Prologue, \\SPX Attestion ext., ephemeral key}};
\postlevel
\mess{edge}{F}{client}
\node[anchor = east](plc) at (mess to) {\shortstack{Prologue \\Server }};
\postlevel
\mess{client}{R}{server}
\node[anchor = east](e) at (mess from) {\shortstack{e}};
\mess{server}{R}{client}
\node[anchor = west](e) at (mess from) {\shortstack{e, dhee, s, dhse}};
\postlevel
\mess{edge}{B}{server}
\node[anchor = east](arb) at (mess from) {\shortstack{SPX \\Attestation \\report}};
\postlevel
\mess{client}{}{server}
\node[anchor = east](bb) at (mess from) {\shortstack{s, dhse}};
\mess{server}{R}{client}
\node[anchor = west](bb) at (mess from) {\shortstack{e, dhee, s, dhse}};
\mess{client}{}{server}
\node[anchor = east](bb) at (mess from) {\shortstack{e, dhse}};
\mess{server}{G}{edge}
\node[anchor = west](bb) at (mess from) {\shortstack{session key}};
\end{sequencediagram}
\caption{Showing the Noise_XX pattern handshake with \sys enabled Noise framework.}
\label{fig:noixe}
\end{figure}

Applications can choose to specify and use their own defined protocol
between their end users and the web service. NoiXe allows edge
functions to support any secure protocol that can be built using the Noise
protocol framework. A Noise protocol starts with prologue exchanged
between a server and client. For example, suppose a server
communicates a list of Noise protocols that it supports to
clients. Applications on the client side will then choose among them
and execute that protocol to communicate with the server. \sys uses
the 
prologue to perform the detect operation on the edge and server side. When a
client tries to connect to a server using a Noise protocol, it is
directed to a NoiXe-enabled edge function. 
%It
The NoiXe edge function leverages the hand shake
pattern information of the protocol being used to choose the appropriate
messages to carry out its other functions. First, it creates a secure
channel with the server using the same protocol as requested by the
client. Then, the edge function and the server carry out the bind operation,
i.e., use the channel to send and receive an attestation report created
in a secure enclave using the ephemeral key shared and verified using certificates. 
On successful mutual attestation, a NoiXe enabled
edge function relays the same channel to the client, while acting as a
Noise proxy for it to carry out its handshake with the server. This
ensures that security expectations of the clients are honored and the
channel that was bound is the one that is getting used for the client-server communication.

On the edge function side, after a detect operation, the 
edge function 
capable of the equivalent NoiXe protocol
initializes a
handshake state object for that protocol and keeps it up to date with the state on the server, while
acting as a forwarding proxy for communication between the web service
and the client. Since Noise uses
symmetrical encryption, there is no need to replicate the execution of
the protocol on the edge function
side. Instead, it selects the appropriate messages from the protocol's handshake message pattern
(part of the Noise handshake state) to obtain the required information from
the server, e.g., current hash state.
Finally, the session keys are granted by the server to the edge function before the last message is exchanged
between the server and client as part of another attestation report to prevent eavesdropping or impersonation.
In the cases where the last message in the handshake pattern is from
the client side, the server grants the
session keys after it finishes the crypto function. The handshake is only carried out
whenever an attestation request can be satisfied (indicated by attestation extension).
The handshake is aborted if a requested attestation is not received or is invalid, or the Noise
session information does not match the signature in the attestation.

\autoref{fig:noixe} shows the handshake patterns for Noise_XX between an initiator (client) and
responder (server), along with the SPX operations marked with same letters as defined in ~\autoref{sec:generic}. 
Noise_XX is used to start a compound protocol called Noise pipe~\cite{whatsapp}.
Noise pipes do not assume any previous communication between a client and server.
The Noise_XX handshake pattern supports mutual authentication and transmission of static public keys
that are stored and used as pre-messages in the next Noise_IK pattern to create the session.
Important to note is that binding happens only after the DH function is performed.
Identifying this step requires knowledge about the protocol. NoiXe
achieves this by the replicated state and performing the attestation after first DH operation
is performed, thus preserving the protocol handshake pattern as well as avoiding MITM attacks. 
Other protocols are supported in similar manner.

%\ada{Is there a way to shade, or otherwise mark in the two figures,
%  parts of the exchange that correspond to different SPX operations
%  (relay, grant?... only bind is shown) We need to somehow highlight
%  them better, and then show them as the common part of both cases. Also, TLS
%people are familiar with. Does it make sense to include a pseudo
%example. }
\section{Implementation}
\label{sec:impl}

We implemented \sys prototypes for Ubuntu 14.04 using the Intel SGX Linux SDK. We highlight interesting
implementation details for TLS and Noise protocols below. 

{\bf TLX.}
We prototyped TLX by modifying the TLS implementation on top of the
mbedTLS libraries. We leveraged the extension 
feature introduced in TLS 1.2 to integrate attestation request and
verification into the handshake. 
We then ported the modified library into an enclave to support SPX functionality.

{\bf NoiXe.} We modified the Noise protocol framework to support a new role of `handler' to create an
intelligent Noise edge function that allows replication of the crypto
  state %-- somewhere in the background on Noise this needs to be clear
  %that it's needed and what it corresponds to. }
of the Noise protocols. To run a Noise
protocols in an SGX enclave, we ported the full Noise protocol
framework to run inside an enclave, 
including all its crypto functions. Using it, we implemented a proxy
that runs as the edge function 
and a server that uses the \sys enabled Noise framework - NoiXe. We
had to disable use of vectorization 
in blake2 and ChaChaPoly because SSE2 optimizations are not supported in an SGX enclave with 6th
generation Intel processors.

{\bf Remote Attestation for Enclave.}
Remote attestation in Intel SGX is designed to use trusted servers,
i.e., Intel servers or other parties 
that are granted the processor secret keys under suitable agreements, to verify the attestation
measurements. However, in our implementation, we bypassed the use of those servers to simplify our
implementation. This does not lead to any loss of generality of our prototype or evaluations because
we posit that parties deploying edge functions would act as those
trusted entities as per Intel. %\ada{has this been done by others, if
  %so add Similar simplification has been used in the implementation of
  %other SGX-based solutions. }
  %\ketan{not sure and no one mentions it ... at least we have mentioned it.}
\section{Evaluation}
\label{sec:eval}
An \sys enabled protocol preserves $E^3$ semantics while allowing edge function full access to the web traffic, with affordable impact on perceived latencies. The evaluations demonstrate that the design and implementation of \sys achieve this goal. 
%which is the functional main goal of this work.

\subsection{Security Analysis}
\label{sec:securityanalysis}
\noindent {\bf $E^3$ security semantics  $\cong$ end-to-end security.}
Using \sys, an edge function gains access to the encrypted traffic but only within a secure SEE enclave. With hardware provided protection, it remains secure from any other software component in the system and any communication 
%from SEE outside it 
from outside the SEE
is secured by the protocol, same as in the default two-party case. 
%By not exposing traffic to any other software component in the path except the edge function, the client-server connection remains logically end-to-end secure. 
In that sense, we claim that $E^3$ security semantics is equivalent to the conventional end-to-end security afforded by encryption using a secure protocol.

\PP{Preserved security properties of protocols} \sys ensures that the security properties of the original protocol for which it is developed remain preserved. For instance, for TLX, it ensures that the connection between client-edge-server remains same as a client-server connection.
%\ada{same -- meaning? same as?}. 
First, the identities of all parties can be authenticated, thanks to the bind step between the edge and server, while the client-server are authenticated using the TLS record protocol. Second, the channel attested is the channel that gets used by the edge function to interact with the client, thanks to the relay step. Finally, the grant before the handshake finishes, from within the SEE, ensures that the shared (negotiated) secret remains secure from any other software entity, including privileged software at the edge, ensuring the reliability of the handshake. Similar discussion is valid for any of the  Noise-based protocols and their corresponding properties. By preserving security properties, a \sys enabled edge remains secure from attacks that those protocols protect against, including lago attacks based on introspection of memory used by applications using these protocols. %, thanks to the use of an OS agnostic SEE. 
However, these protocols are not designed to protect against lago attacks that may use impersonation as they assume end points to be secure.

\PP{Protection from impersonation by privileged attackers} \sys prevents lago attacks such as the cuckoo and TOCTTOU attacks by using ephemeral keys during the bind phase and ensuring that the grant is done before the handshake negotiation finishes. By doing this it ensures that no two sessions have the same ephemeral key and a malicious attacker cannot wait for the end of the protocol and try to get access to the shared private key. Further, the sharing remains within the SEE and is never exposed outside of it. These, combined, prevent any privileged attacker from impersonating as a  web service to an edge function, or vice versa, to gain access to the session key. 

\PP{Comparison with relevant alternatives}
In comparison to recent proposals with similar goals, the \sys approach differs in the following ways.
{\em SafeBricks} proposes to use Intel SGX and controlled development of NFs to protect them against Iago attacks~\cite{iago} when deployed in the cloud.
It relies on IPsec isolation for protection of the traffic. Use of IPsec can be unreasonable in the target use cases for edge computing. For use cases such as IoT
data aggregation, and caches for arbitrary clients, end devices may not support IPsec. Further, it remains susceptible to the TOCTTOU and cuckoo 
attacks as it exchanges secrets after the connection between the client and NF with SEE is established. {\em mbTLS} proposes to modify TLS in a similar 
way as proposed by \sys, but is limited to TLS itself. Although, sufficient for use in edge computing with TLS, it cannot be applied if some other protocol is
used. In comparison, \sys achieves $E^3$ security semantics for existing protocols while remaining completely transparent to end user devices making it well suited 
for edge computing. 
However, \sys does have impact on performance of those protocols, which we present next.

\subsection{Performance}
\label{sec:evalperformance}
\sys ensures that edge functions remain transparent for clients, which is critical in edge computing for practical deployability. In this section, we evaluate the performance overhead incurred to carry out the additional steps proposed by \sys.

\begin{table}[t]
\begin{tabular}{p{1cm}| p{6cm}}
\hline
{\footnotesize\bf Setup} & {\footnotesize\bf Configuration}\\ 
 \hline
 {\footnotesize Client} &  {\footnotesize Intel(R) Xeon(R) CPU E5-1620 v2 @ 3.70GHz, 16GB DIMM DDR3 Synchronous RAM, Intel Ethernet Connection (2) I219-V}\\  
  \hline
{\footnotesize Edge} &  {\footnotesize Intel(R) Core(TM) i7-6700 CPU @ 3.40GHz, 16GB DIMM DDR3 Synchronous RAM, 128 MB configured for SGX, Intel Ethernet Connection (2) I219-V }\\  
 \hline
{\footnotesize Server }&  {\footnotesize Intel(R) Core(TM) i7-6700 CPU @ 3.40GHz, 16GB DIMM DDR3 Synchronous RAM, 128 MB configured for SGX, Intel Ethernet Connection (2) I219-V}\\    
 \hline
{\footnotesize Network} &  {\footnotesize LAN over 512 Mbps connection. RTT (ms) Client-Proxy: 0.964, Proxy-Server: 0.901, Client-Server: 0.962}\\  
 \hline
 \end{tabular}
\caption{Experimental setup.}
\label{tab:setup}
\end{table}

\PP{Experimental Setup}
%We start with describing out experimental setup. 
\autoref{tab:setup} lists the experimental setup used in the experiments. 
We used machines with 6th generation Intel processors with SGX support for the edge functions and web services.
We implemented a TCP client, proxy (edge function), and server, which communicate using TLS (mbedTLS library)
and Noise protocols using a standard framework as our baselines. Then, we added \sys capability to the edge
and server side, i.e., implemented the portion that handles the encryption state (proxy and server both) to
run in an SGX enclave, and used TLX (modified mbedTLS library) and NoiXe (modified Noise framework libraries).
We used 128 bit AES encryption in RSA for TLS with a certificate size 3KB, and ChaChaPoly cipher,
256 SHA2, and 56 bit curve488 key for the Noise based protocols. In both cases, we used a 512 byte attestation
report generated by the processor inside an SGX enclave. We used an echo protocol 
%\ada{benchmark}
%\ketan{echo protocol is used in networking world for performance ... you can implement different benchmarks using it}
over a TCP socket connection to
carry out our experiments. We intentionally used very fast network between the components to be able
to accurately highlight the overhead associated with \sys especially free of the network delays
which may amortize the effect of some of the overhead. 
In that sense, the experimental setup demonstrates clearly the performance impact on latency due 
to \sys given that impact of the RTTs remains constant.
%\ada{{. -- this is not entirely true. part of the performance impact is from the RTT between the edge and the server, no? remove this statement?}

\PP{Workload \& Measurements} We use application kernels that are representative of common processing steps in edge functions 
as our workload. 
Further, 
our measurements are focussed on latency, i.e., the time as seen from unmodified clients which is different than earlier
work which primarily focussed on throughput due to their target use cases~\cite{mbtls}. We also report the CPU overhead of \sys 
on the edge computing infrastructure, which is important to highlight given that resources at edge infrastructure
may be limited compared to well provisioned cloud or semi-fixed function middleboxes.

\begin{figure}[t]
\centering
\subfigure[]{\label{fig:tlstx}\includegraphics[width=0.45\columnwidth]{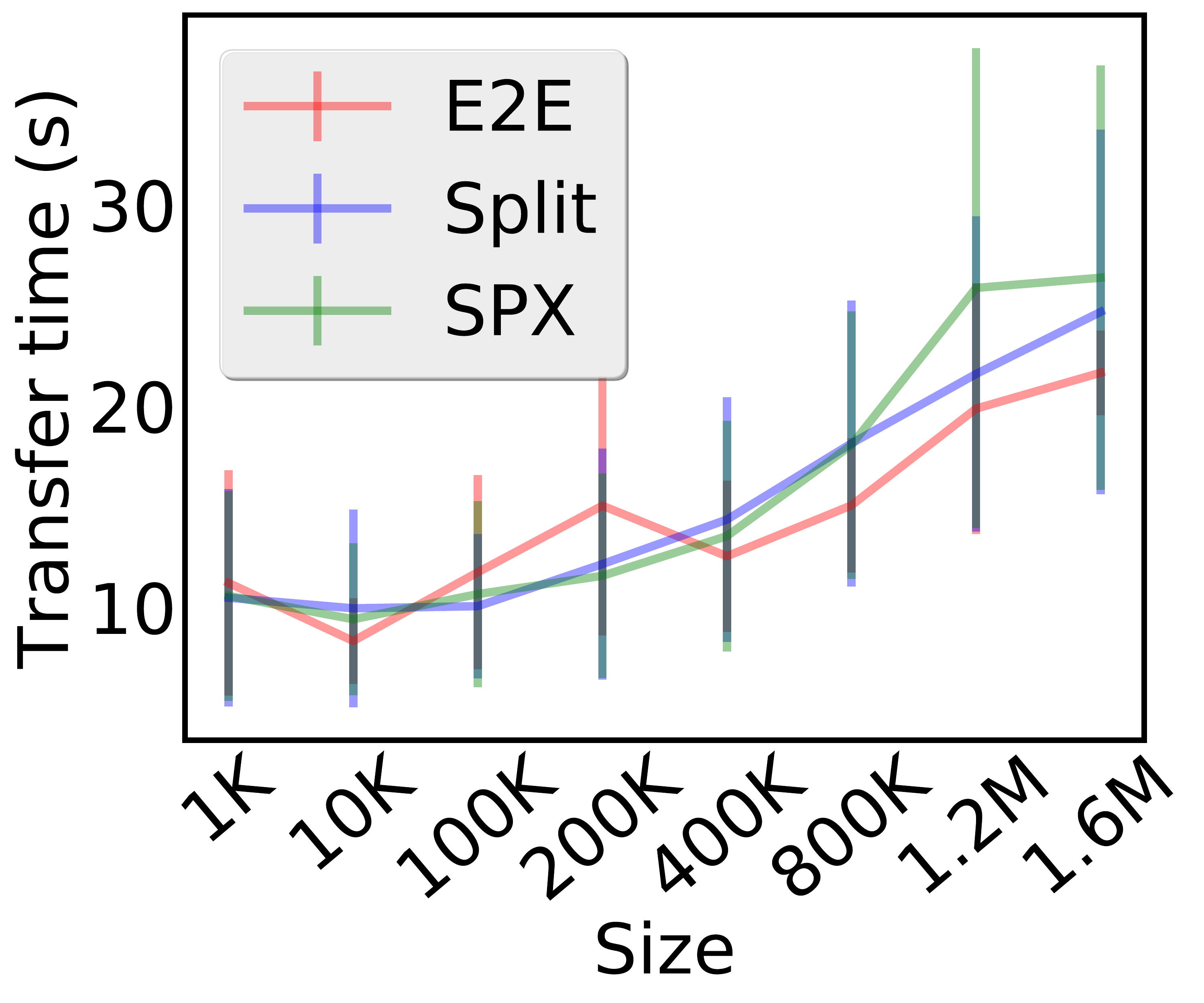}}
\subfigure[]{\label{fig:webload}\includegraphics[width=0.45\columnwidth]{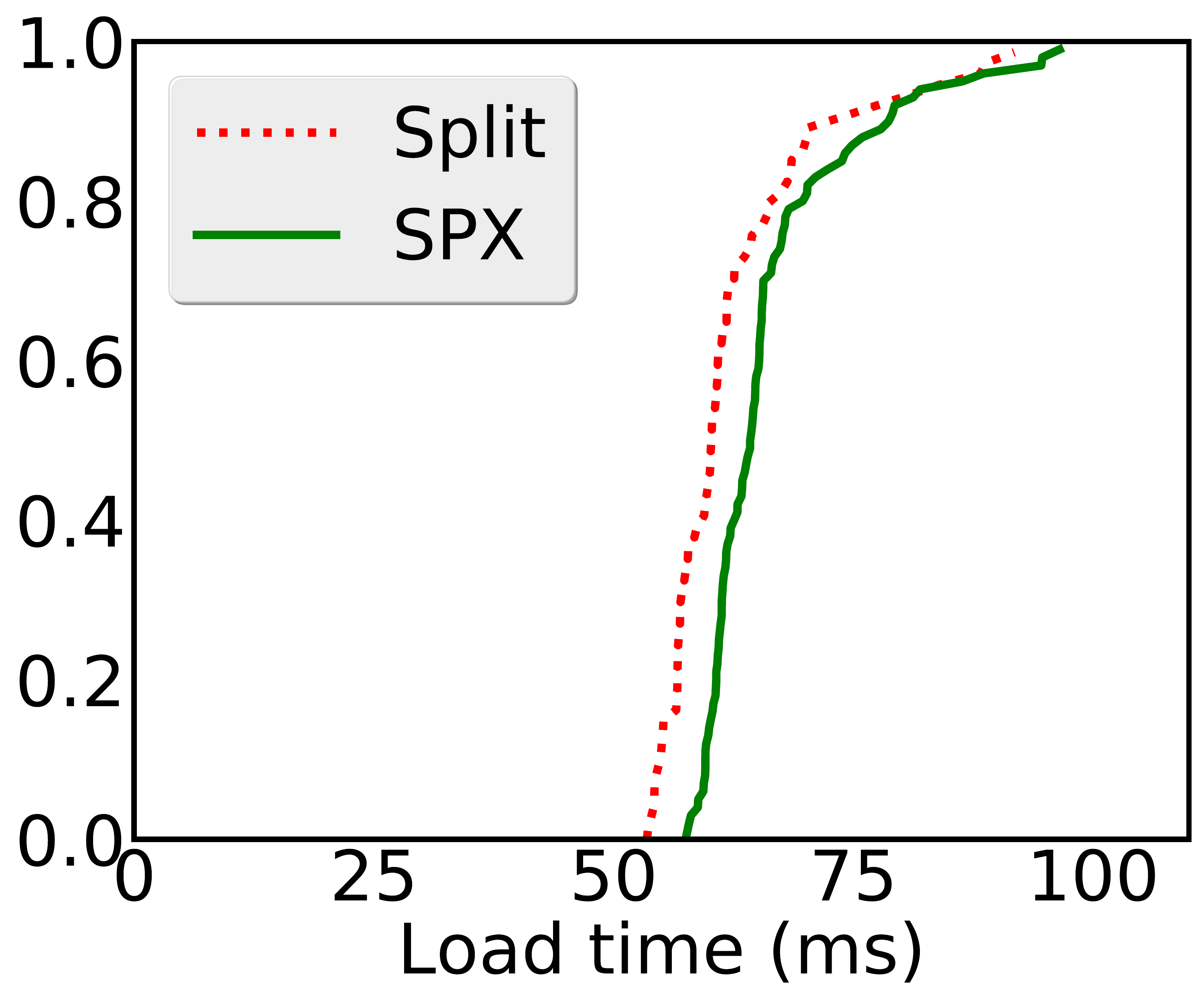}}
\caption{(a) Showing time measured on client side in file transfer using TLX (\sys enabled TLS) over TCP. (b) Showing CDF of loading time for locally replayed web page load times for Alexa's top 100 web pages over TLX.}
\end{figure}

\begin{figure}[t]
\centering     %%% not \center
\subfigure[]{\label{fig:b}\includegraphics[width=0.45\columnwidth]{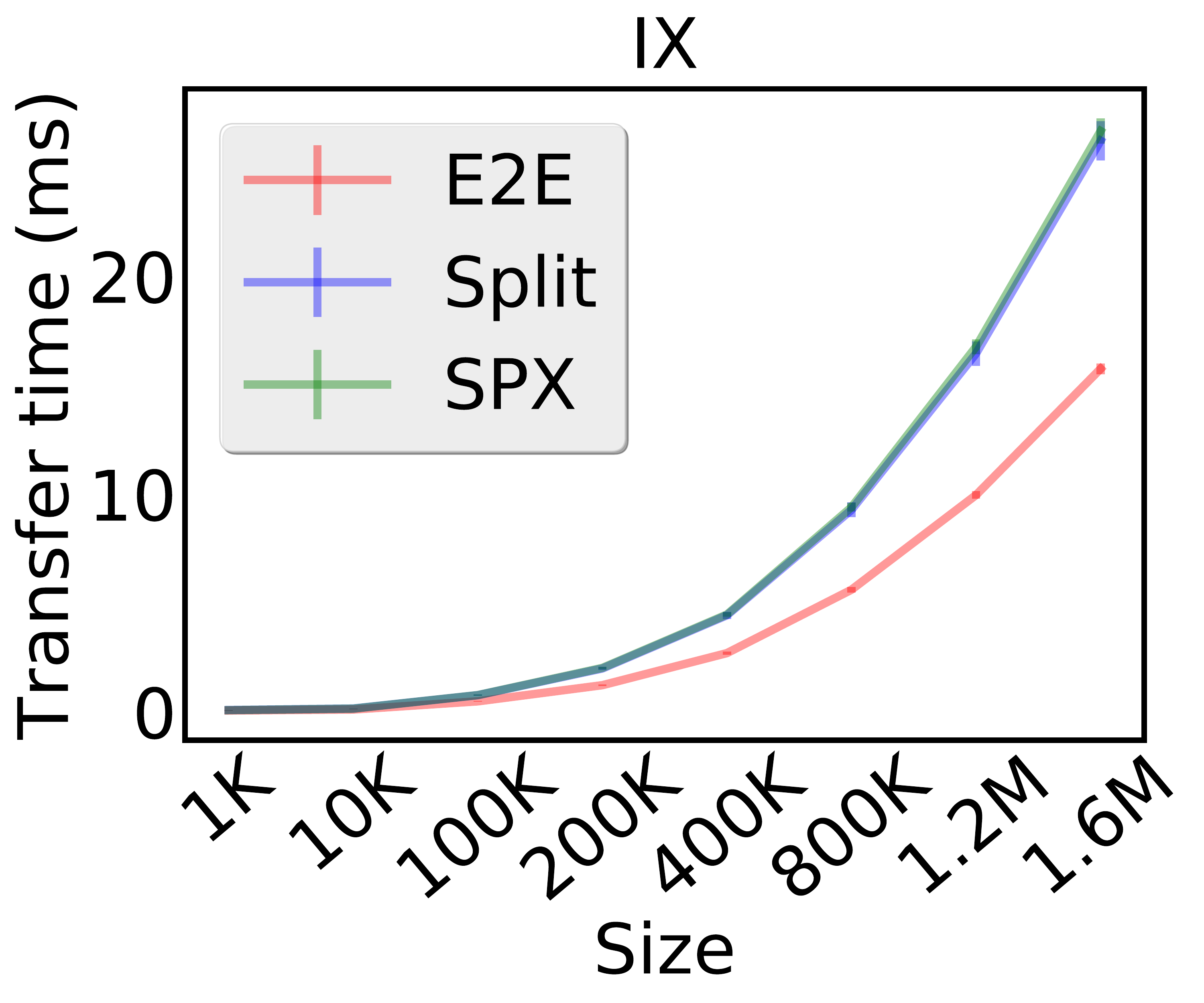}}
\subfigure[]{\label{fig:b}\includegraphics[width=0.45\columnwidth]{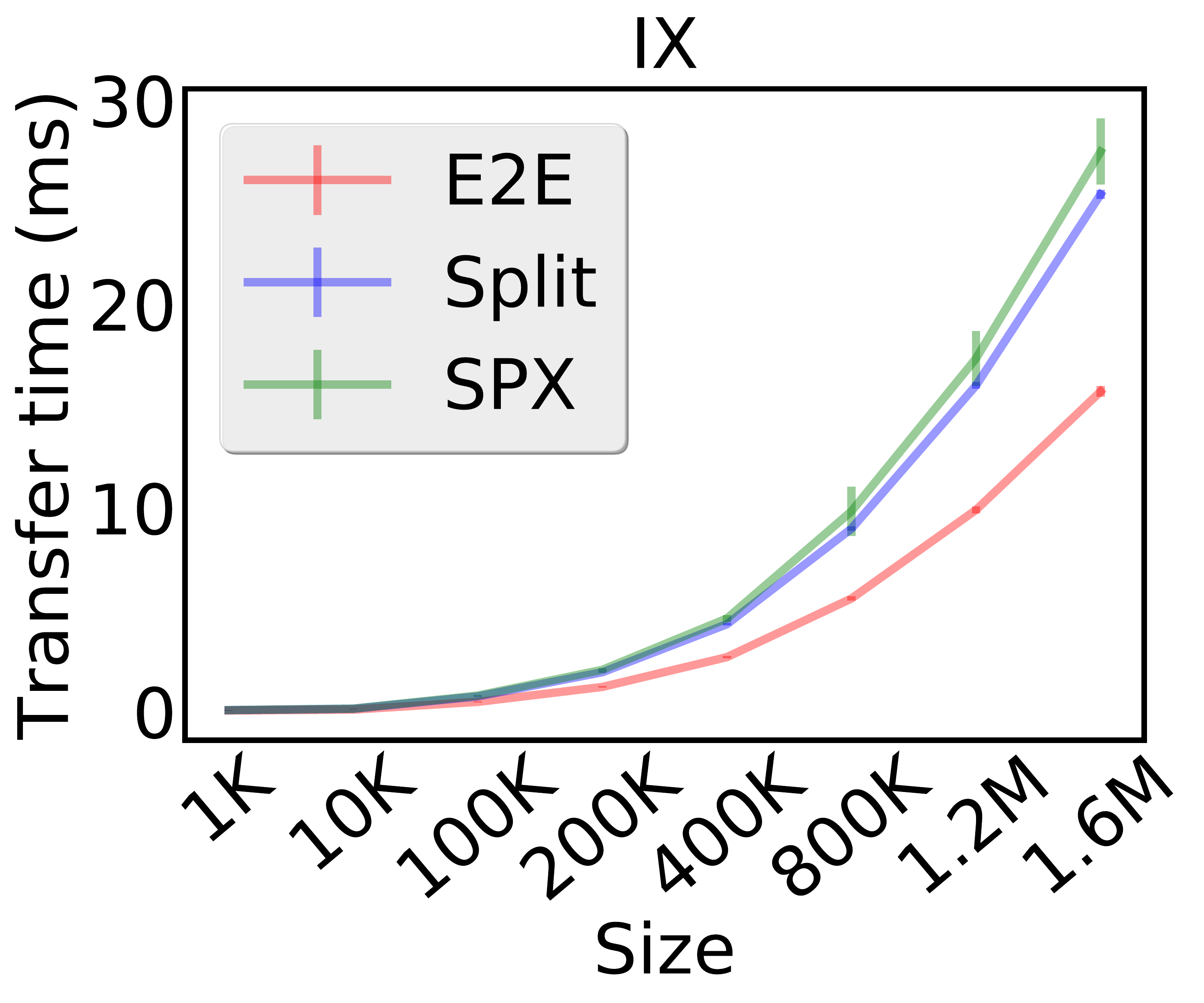}}
\subfigure[]{\label{fig:b}\includegraphics[width=0.45\columnwidth]{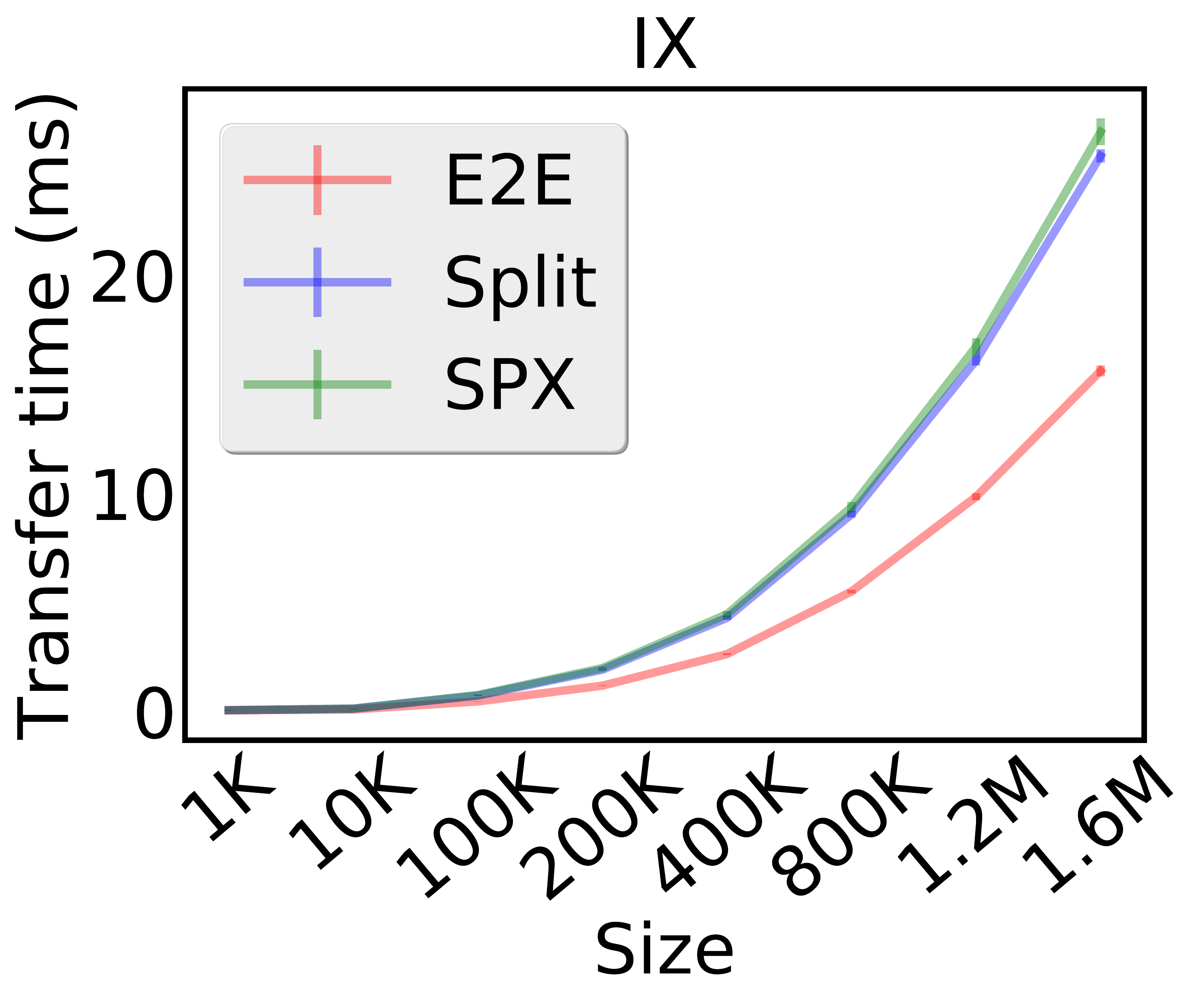}}
\subfigure[]{\label{fig:b}\includegraphics[width=0.45\columnwidth]{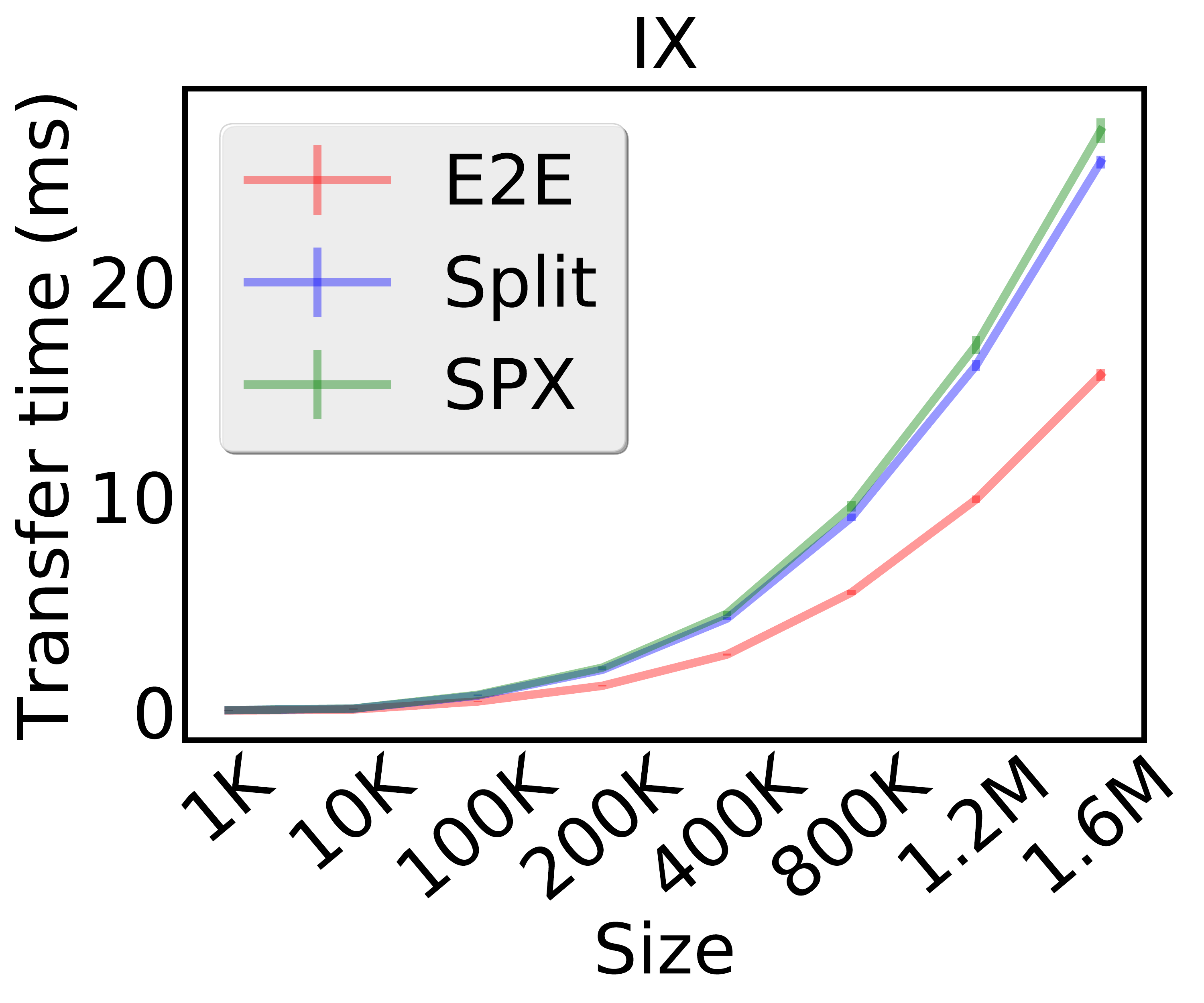}}
\subfigure[]{\label{fig:b}\includegraphics[width=0.45\columnwidth]{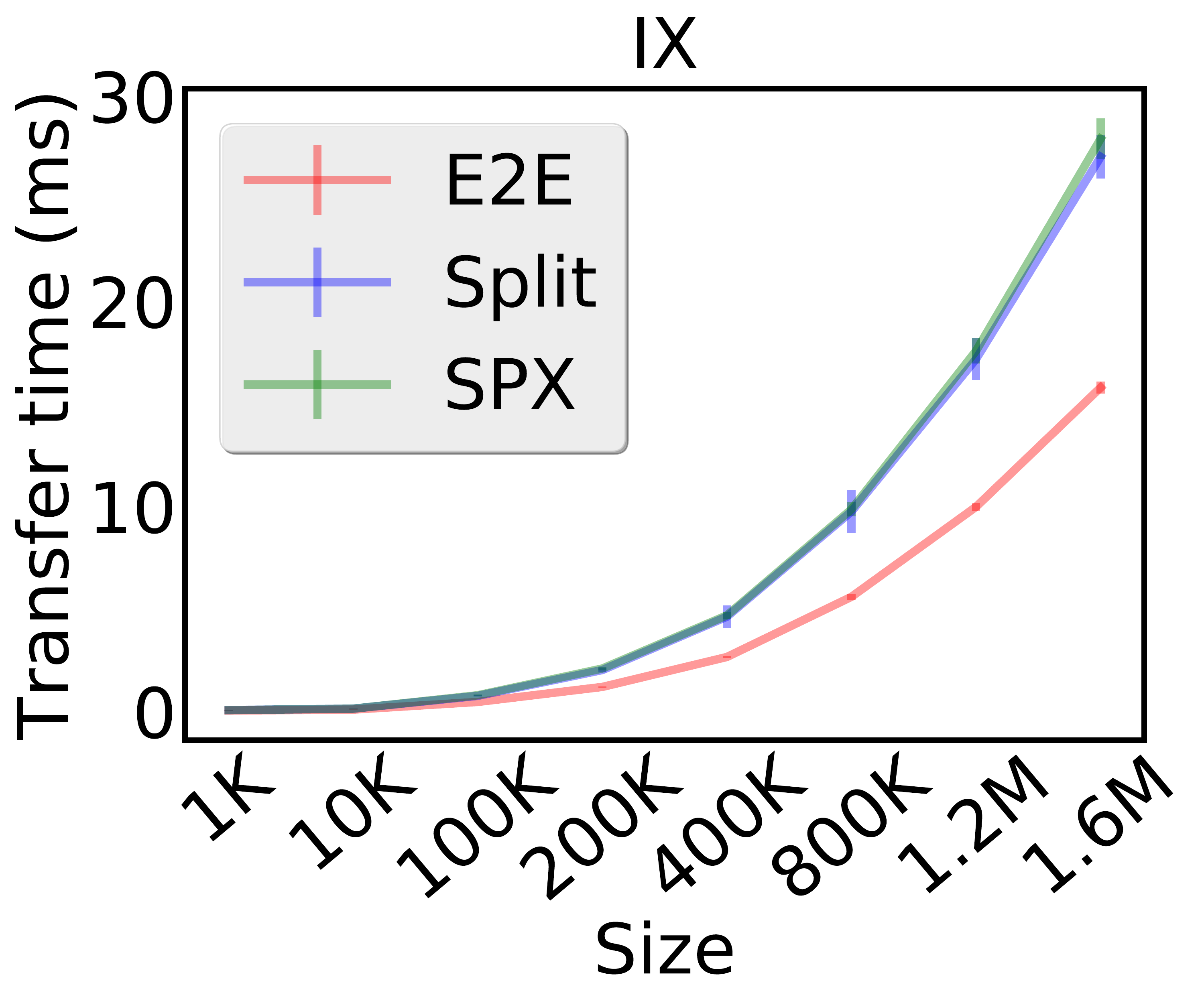}}
\subfigure[]{\label{fig:b}\includegraphics[width=0.45\columnwidth]{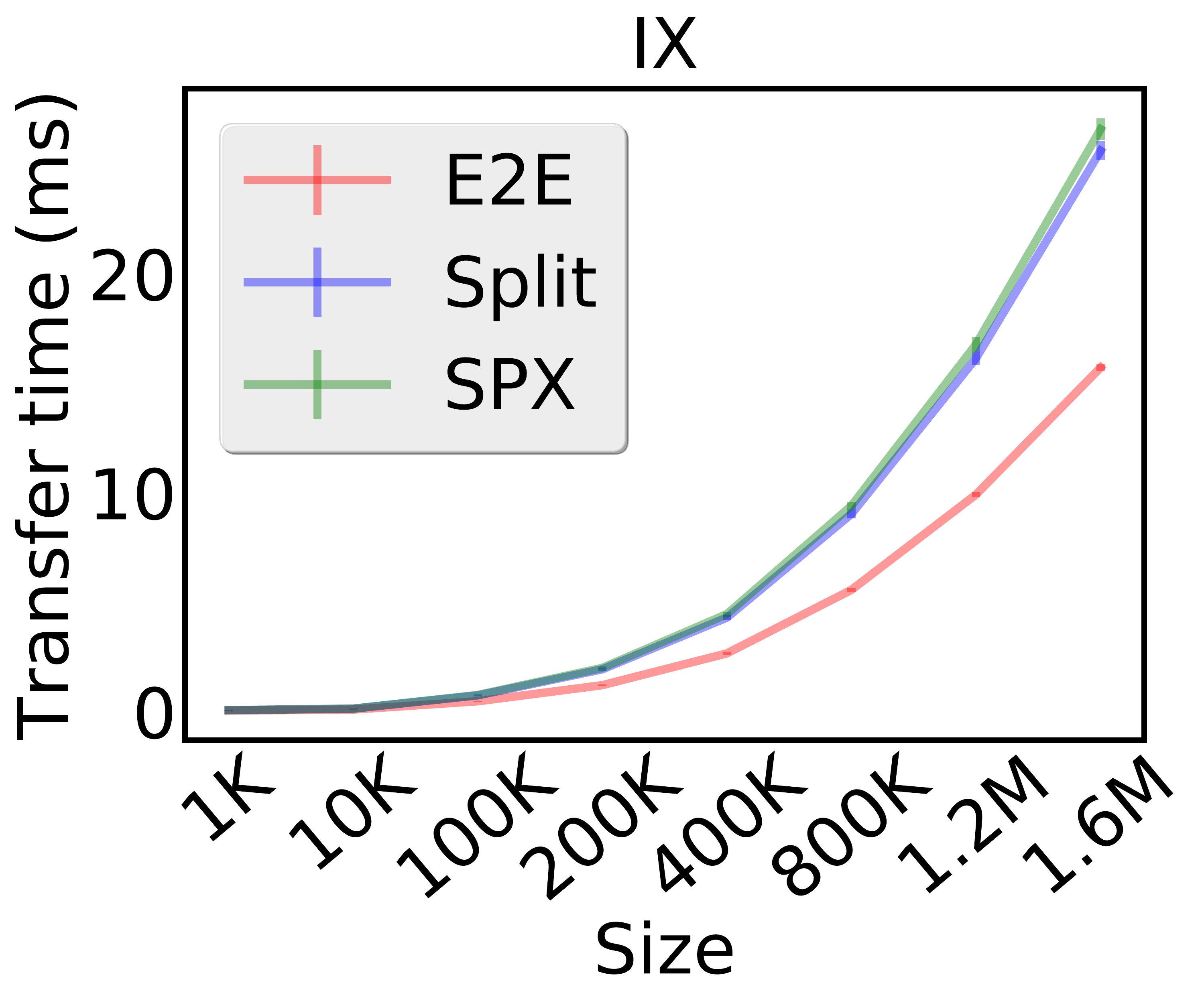}}
\caption{Showing time measured on client side in file transfer using different handshake patterns provided by NoiXE (\sys enabled Noise framework) using echo protocol over TCP.}
\label{fig:noisetx}
\end{figure}

\PP{File transfer} File transfer is representative of multiple use cases such as
content caching or IoT sensor data aggregation. Each point in the \autoref{fig:tlstx} and \autoref{fig:noisetx} 
can be used to characterize a different class of edge functions. For instance, on the lower end 1K file size transfer can
be mapped to IoT sensor values where as 1.6 MB file transfer can be thought of as video streaming. 

\autoref{fig:tlstx} and \autoref{fig:noisetx} compare the time taken to transfer files of different sizes. We compare \sys with two baselines corresponding to a direct client-server case without any edge processing (E2E), and a second one corresponding to Split TLS~\cite{sslsplit}. The other solutions mentioned in \autoref{sec:securityanalysis} are not publicly available so we could not include them in the direct performance measurements.
%E2E corresponds to a baseline where no 
%edge functions are used, and Split corre 
The experimental results shown are averaged over 20 runs along with one standard deviation.
\sys incurs modest overhead of 12-15\% over the insecure Split proxy baseline. 
This can be attributed to execution inside an SGX enclave, partly due to inherent SGX overhead, and partly due to the copying of 
encrypted network packets from host to enclave. 
%To put it in context with other solutions developed for middlebox introspection, 
The overhead is consistent with %\ada{what's consistent, the overheads or the deviation as a function of block size?}
the reported results in other solutions for introspection of encrypted traffic in middleboxes~\cite{mbtls, safebricks}. 
In addition,  the \sys approach also does not mandate apriori knowledge of what to look for in the encrypted traffic, compared to earlier approaches limited in functionality or with high performance overhead, such as~\cite{mctls,blindbox}.
The higher standard deviation in TLS vs.~Noise is due to the packet size 
used in experiments. TLS experiments used blocks of 1KB for the full file transfer where as Noise mandates the use of messages to be less than or equal to 65535 bytes by design for simplicity in testing, reducing the likelihood of memory handling or integer overflow related issues and efficient implementation of stream ciphers.

\PP{Web page loading}
To understand how  \sys would perform in realistic settings where the communication may be over multiple connections and not always of the same size objects, we mimicked the web page loading process with \sys. We recorded and replayed Alexa's top 100 websites 
using the Mahimahi tool~\cite{mahimahi} via a \sys-enable proxy. \autoref{fig:webload} shows the cumulative distribution function (CDF) 
of the replayed webpage load times with proxy only and with \sys-enabled proxy. Evident from the graph is that there is modest 
overhead from \sys in full web page loading. We attribute this to the structure of the web pages which include a large number of small objects, 
where the overhead of \sys is not amortized. However, it is important to consider the fact the we are locally 
replaying web pages over a very fast networks. The web page loading, which is usually in order of seconds finishes in tens of milliseconds, 
thus presenting the worst case for TLX. This is also evident on the right side in the figure: as the time to load increases, 
the performance gap diminishes.

\begin{figure}[t]
\centering     %%% not \center
\subfigure[]{\label{fig:hs-a}\includegraphics[width=0.49\columnwidth]{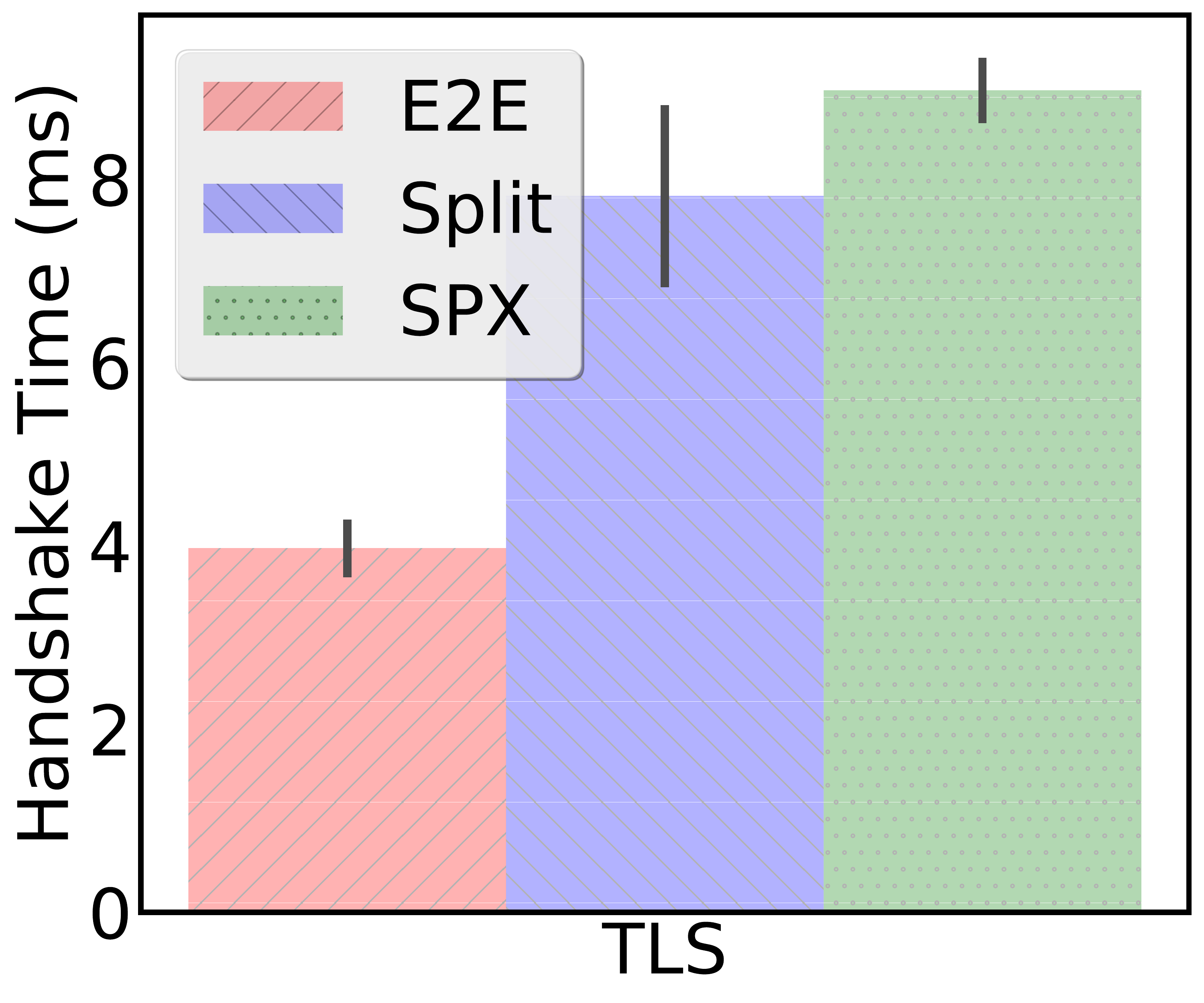}}
\subfigure[]{\label{fig:hs-b}\includegraphics[width=0.49\columnwidth]{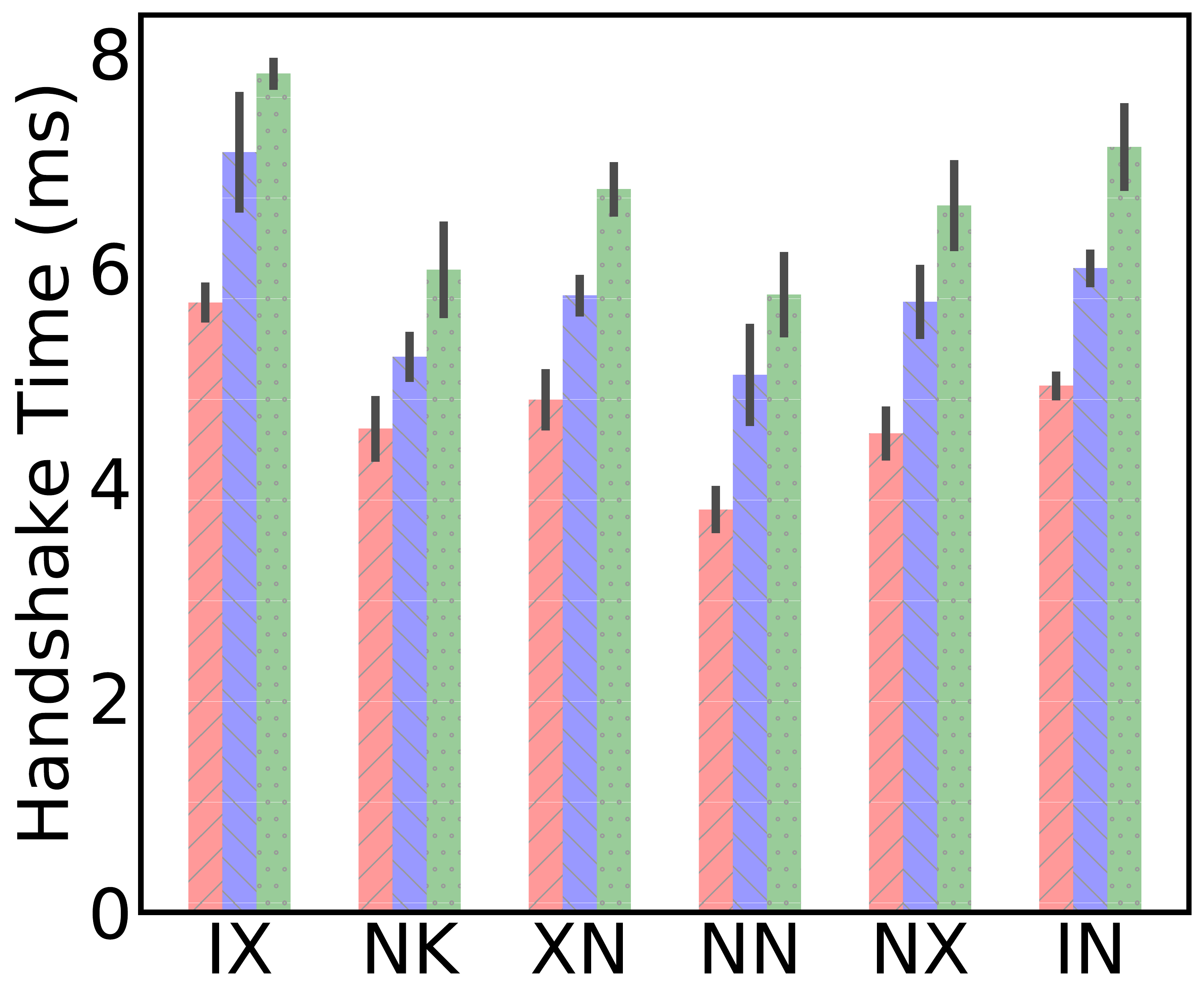}}
\caption{Comparing the handshake time of \sys enabled protocols (TLX, NoiXe) as observed on client side.}
\label{fig:handshake}
\end{figure}

\PP{Handshake} \autoref{fig:handshake} shows the time taken to carry out handshakes
when a client connects to a server without any proxy (E2E), with a Split connection proxy, and finally, via
an \sys-enabled proxy and server for TLS (\autoref{fig:hs-a}) and 5 interactive handshake patterns (\autoref{fig:hs-b}). Readers are directed to the 
Noise specifications~\cite{noise} for details about the handshake patterns in Noise and its description terminology. %\ada{should we prepare a ``brief'' on Noise and include as supporting documentation?}
%\ketan{I think specification is fairly easy to understand and is available online}
The results shown are averaged over 20 runs to show one standard deviation. We observed an overhead of only 12\%-15\% 
over the baseline corresponding to the Split approach, which is the only available approach to enable edge functionality. 
%\sys outperforms the only solution reported in literature~\cite{blindbox} that offers such 
%security semantics.
%\ada{However, in offering such security semantics that work is limited in that it only allows a limited level
%of access to middle boxes. -- I removed the blindbox comparison. I don't understand this sentence}. 
%\ketan{I missed this sentence.}

\PP{Extra bytes and RTTs on wire during handshake}
In \sys, extra messages are exchanged between the edge and server during setup of up a secure connection. 
For TLS we only add one extra message between edge and server corresponding to 1 extra RTT and for
noise based protocols extra RTTs are 1 or 2 depending on whether the last message is from server or client
respectively.
%while ensuring that end clients do not see any change. 
The number of extra bytes exchanged between the edge and
server due to \sys-enabled edge functions is $2*Size_{attestation-report} + Size_{session-key}$ also listed in 
~\autoref{tab:overhead}.
This demonstrates that the overhead of \sys in terms of extra bytes is reasonable, it is dependent on the protocol itself, and \sys is suitable for
short-or long-lived connections. %, and is dependent on the protocol itself.

\begin{table}[t]
\begin{tabular}{p{2.2cm}|p{2.2cm}p{2.2cm}}
\hline
{\footnotesize Protocol} & {\footnotesize Extra bytes} & {\footnotesize Extra RTTs}\\
\hline
{\footnotesize TLX} & {\footnotesize 1152} & {\footnotesize 1}\\
{\footnotesize NoiXe} & {\footnotesize 1090} & {\footnotesize 1 or 2}\\
\hline
\end{tabular}
\caption{Overhead due to SPX in terms of extra bytes and RTTs.}
\label{tab:overhead}
\end{table}

%\ada{this has to be more precise, some real numbers in terms of B and RTTs per connection. are there btw same number of RTTs in mbtls? would be good if we had less. }

Due to space limitations, and the current prevalence of TLS, we only show results for TLS for the remaining benchmarks. 
However, we report that we observed similar trends for the Noise protocols as well.

%\begin{figure}[t]
%\centering
%\includegraphics[width=\columnwidth]{fig/mulcon}
%\caption{showing variation of time measured in file transfer using TLX with number of concurrent connections.}
%\label{fig:connnections}
%\end{figure}

\begin{figure}[t]
\centering     %%% not \center
%\subfigure[]{\label{fig:a}\includegraphics[width=0.45\columnwidth]{fig/edgecpu}}
\subfigure[]{\label{fig:connnections}\includegraphics[width=0.45\columnwidth]{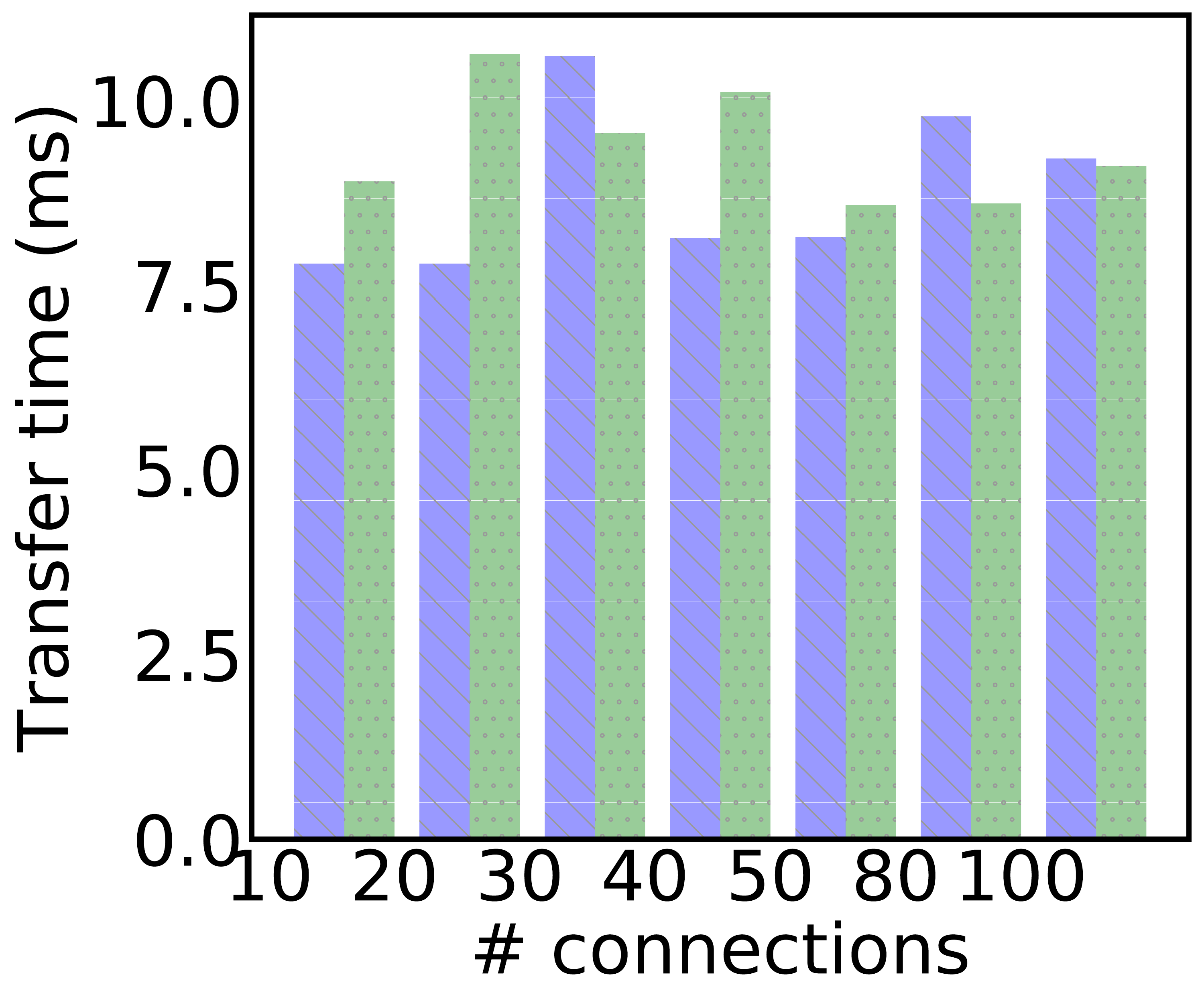}}
\subfigure[]{\label{fig:b}\includegraphics[width=0.45\columnwidth]{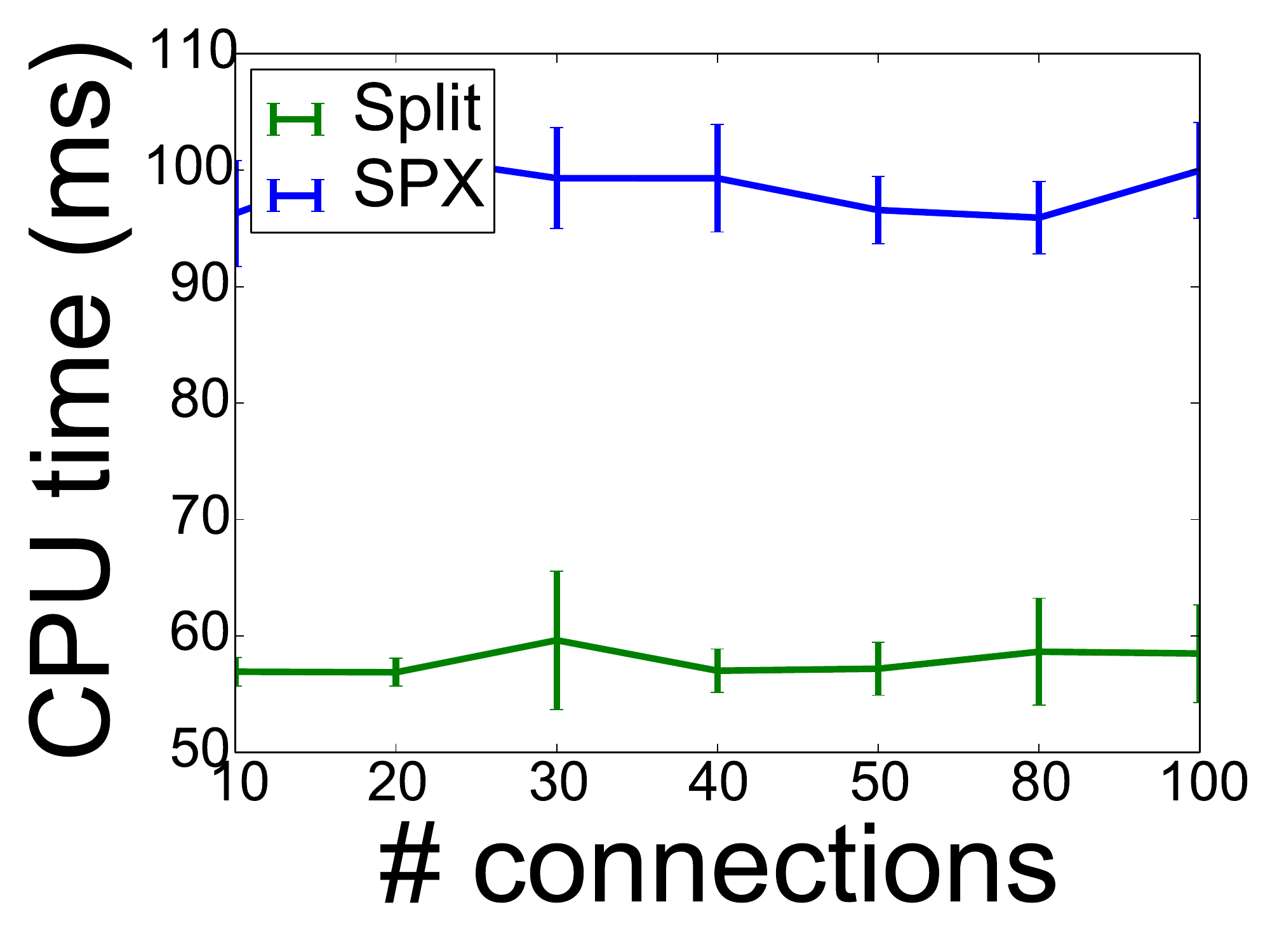}}
\caption{(a) Showing variation of time measured in file transfer using TLX with number of concurrent connections (b) Comparing CPU time spent on proxy side measured file transfer using concurrent.}
\label{fig:cpu}
\end{figure}

\PP{Scalability with concurrent connections}
\autoref{fig:connnections} shows the handshake time measured on the  client side for diffrent number of concurrent connections. 
Evident from the figure is that the performance of \sys is not constrained by the number of concurrent connections.
This is important for \sys to be a solution in edge computing scenarios because the edge infrastructure is by 
definition resource constraint. Further, it also highlights another point: with \sys, edge functions can 
securely service multiple clients without being limited by the amount of memory addressable by SGX enclaves.

\PP{Scalability of edge functions}
We report that \sys puts substantial overhead of nearly \textasciitilde2x in terms of CPU time during its operation. This has not been 
reported in earlier works that utilize SEE to develop introspection solutions for middleboxes.  
The consumption is due to memory copy of network packets from host to enclave and overhead due to memory page encryption and lack of vector instructions for cryptographic operations.
This is substantial on its own. However, compared to BlindBox~\cite{blindbox}, which uses much more CPU-intensive searchable 
encryption while supporting only reduced functionality at the middleboxes with similar security guarantees, \sys provides 
full access to encrypted traffic with comparatively lower overheads. It may not be suitable for middleboxes which are simply 
preprovisioned boxes, but is especially important for edge computing, because this implies that developers have full 
flexibility to implement any logic for their edge functions and required compute capacity can be provisioned accordingly.

\section{Related Work}
\label{sec:related}

In addition to the solutions discussed in the earlier sections~\cite{safebricks,mbtls,mctls,blindbox,gentry,sslsplit}, \sys builds on other prior work related to secure execution environments.

\noindent {\bf Edge function security.}
%\ada{Research has drawn less attention than edge computing research. ?} 
Although several surveys~\cite{sen2013security,openi2013security,suo2013security,roman2016mobile}
have pointed out security and privacy challenges for edge functions,
they do not handle cases of compromised operating systems but instead, focus on securing
an execution environment.
In an attempt to address the handling of encrypted traffic in edge services,
one study~\cite{airbox} proposed Airbox, which is based on Intel SGX. Airbox, however,
is not protocol-centric, nor does it address the limitations of the SEE hardware or prevent cuckoo or TOCTTU attacks. \sys fills this gap.

\noindent {\bf Secure communication channel.} Assuming a similar threat model in which a
host is untrusted, some studies have focused on
building a secure communication channel on the same machine. Zhou et al.~\cite{zhou:building}
outline an approach to secure the data-transfer between a user's I/O device and a program
trusted by the user. Jang et al.~\cite{jang:secret} propose SeCReT, a framework for building
a secure channel between SEE and non-SEE. 
%\ada{since the start of the paper referenced SEE, I've been changing the SEEs to SEEs, but they keep coming up... }
Unlike the approaches in these studies,
\sys targets the establishment of secure end-to-end channel. EndBox proposes to run middlebox functions on client side using Intel SGX by leveraging VPN connections between clients and enterprise VPS servers. The approach is not applicable in edge computing as the edge functions run on edge infrastructure and may not be suitable to run on individual client devices. A simple example is a web caching edge function which is already prevalent in client browsers, but using caching across clients has its own set of security issues. Further, from a performance perspective, they may not offer acceptable performance which is the core reason to deploy edge functions.

\noindent {\bf Use of SEE-based remote attestation.} 
%\ada{can you prune this paragraph a little, too long... }
Several systems built on top of shielded execution environments~\cite{aws:hsm,chen:overshadow,hofmann:inktag,ta:splitting,yang:using,zhang:cloudvisor,raj2011credo,baumann:haven,owusu:oasis,schuster:vc3} 
 focus on providing a secure execution environment. However, they do not provide for establishing the trustworthiness 
of the communication channels with such primitives, which is an inseparable and critical part of edge 
computing. Several studies~\cite{mccune:flicker,trustvisor,martignoni:cloud-terminal} have used remote attestation to provide a remote party trustworthiness of the local TCB. Haven~\cite{baumann:haven} leverages Intel SGX to prevent code and data from tampering
by an untrusted cloud provider. 
Although the above-mentioned studies are concrete examples of using hardware-assisted remote attestation,
they either fail to provide details about how to perform remote attestation in a communication protocol
or assume only the naive protocols. \sys fills this gap in the protocol.
Another approach replaces the TLS certificate by the attestation
identity key (AIK) certificate. However, this solution is not suitable for the multi-tenant
virtualized edge-computing scenario because (i) the AIK certificate is typically unique
to one host machine, so it is vulnerable to MITM attack; (ii) the associated costs of
assigning a TLS certificate to each edge function are high;
and (iii) high SEE overhead in decrypting the pre-master
secret and generating the attestation. In comparison, the 
\sys method of binding TLS session with remote attestation
on the server side addresses those concerns
\section{Conclusion}
\label{sec:conc}
This paper presented a new framework -- \sys -- to create secure
protocol extensions that enable edge functions to operate on encrypted traffic, without
compromising end-to-end security properties or edge-related
performance benefits. We coined $E^3$ security properties as
equivalent to end-to-end security properties in edge-computing scenarios.
We demonstrated the feasibility and performance of \sys by prototyping it for
TLS and Noise-based protocols based on Intel SGX. In addition to
providing the desired performance and security properties, 
 \sys addresses difficult problems such as interoperability and
deployability, to ensure that we place minimum constraints on the
developers in the 
nascent edge-computing ecosystem.

\balance
\bibliographystyle{abbrvnat}
\footnotesize
\setlength{\bibsep}{3pt}
\bibliography{spx,p,sslab,conf}
\end{document}